\documentclass{aastex63}

\usepackage{color}

\newcommand{\hhh}{$^{\mathrm{h}}$}
\newcommand{\mmm}{$^{\mathrm{m}}$}

\newcommand{\ddd}{$^{\mathrm{\circ}}$}
\newcommand{\dmm}{$^{\prime}$}
\newcommand{\dss}{$^{\prime\prime}$}

\received{October 20, 2021}
\revised{December 10, 2021}
\accepted{December 15, 2021}
\submitjournal{ApJ}

\shorttitle{The 1.28 GHz MeerKAT Galactic Center Mosaic}
\shortauthors{Heywood et al.}

\graphicspath{{./}{figures/}}

\begin{document}
\title{The 1.28 GHz MeerKAT Galactic Center Mosaic}

\correspondingauthor{Ian Heywood}
\email{ian.heywood@physics.ox.ac.uk}

\author[0000-0001-6864-5057]{I. Heywood}
\affil{Astrophysics, Department of Physics, University of Oxford, Keble Road, Oxford, OX1 3RH, UK}
\affil{Department of Physics and Electronics, Rhodes University, PO Box 94, Makhanda 6140, South Africa}
\affil{South African Radio Astronomy Observatory, 2 Fir Street, Observatory 7925, South Africa}

\author{I.~Rammala}
\affil{Department of Physics and Electronics, Rhodes University, PO Box 94, Makhanda 6140, South Africa}
\affil{South African Radio Astronomy Observatory, 2 Fir Street, Observatory 7925, South Africa}

\author[0000-0002-1873-3718]{F.~Camilo}
\affil{South African Radio Astronomy Observatory, 2 Fir Street, Observatory 7925, South Africa}

\author{W.~D.~Cotton}
\affil{National Radio Astronomy Observatory, 520 Edgemont Road, Charlottesville, VA 22903, USA}
\affil{South African Radio Astronomy Observatory, 2 Fir Street, Observatory 7925, South Africa}

\author{F.~Yusef-Zadeh}
\affil{CIERA and the Department of Physics \& Astronomy, Northwestern University, 2145 Sheridan Road, Evanston, IL 60208, USA}

\author{T.~D.~Abbott}
\affiliation{South African Radio Astronomy Observatory, 2 Fir Street, Observatory 7925, South Africa}

\author{R.~M.~Adam}
\affiliation{South African Radio Astronomy Observatory, 2 Fir Street, Observatory 7925, South Africa}

\author{G.~Adams}
\affiliation{South African Radio Astronomy Observatory, 2 Fir Street, Observatory 7925, South Africa}

\author{M.~A.~Aldera}
\affiliation{Tellumat (Pty) Ltd., 64-74 White Road, Retreat 7945, South Africa}

\author{K.~M.~B.~Asad}
\affiliation{Astronomy Research Group, Independent University, Bangladesh, Bashundhara RA, Dhaka 1229, Bangladesh}

% ...

\begin{abstract}

The inner $\sim$200 pc region of the Galaxy contains a 4 million M$_{\odot}$ supermassive black hole (SMBH), significant quantities of molecular gas, and star formation and cosmic ray energy densities that are roughly two orders of magnitude higher than the corresponding levels in the Galactic disk. At a distance of only 8.2 kpc, the region presents astronomers with a unique opportunity to study a diverse range of energetic astrophysical phenomena, from stellar objects in extreme environments, to the SMBH and star-formation driven feedback processes that are known to influence the evolution of galaxies as a whole. We present a new survey of the Galactic center conducted with the South African MeerKAT radio telescope. Radio imaging offers a view that is unaffected by the large quantities of dust that obscure the region at other wavelengths, and a scene of striking complexity is revealed. We produce total intensity and spectral index mosaics of the region from 20 pointings (144 hours on-target in total), covering 6.5 square degrees with an angular resolution of 4\dss\,at a central frequency of 1.28 GHz. Many new features are revealed for the first time due to a combination of MeerKAT's high sensitivity, exceptional $u,v$-plane coverage, and geographical vantage point. We highlight some initial survey results, including new supernova remnant candidates, many new non-thermal filament complexes, and enhanced views of the Radio Arc Bubble, Sgr A and Sgr B regions. This project is a SARAO public legacy survey, and the image products are made available with this article.

\end{abstract}


\section{Introduction}
\label{sec:intro}

Jansky's discovery that some of his instrumental noise was moving according to sidereal time marked the birth of radio astronomy as a science \citep{jansky1933}. The peak of the extrasolar radio emission was in the constellation of Sagittarius, from the direction of the Galactic center (GC), a finding later confirmed and refined by Reber's multifrequency mapping of the radio sky with a parabolic antenna \citep{reber1944}. In addition to its historic place in the field of radio astronomy, the GC is a region of immense astrophysical interest, and has been studied extensively by many observatories covering all accessible wavelengths.

There are several pieces of evidence that the compact radio source with the designation Sagittarius A$^{*}$ \citep[Sgr A$^{*}$;][]{balick1974,reid1999,reid2004} is coincident with the dynamical center of the Galaxy, and marks the position of a supermassive black hole \citep{genzel1997,ghez1998} with a mass of 4.3~$\times$~10$^{6}$~M$_{\odot}$ \citep{gillessen2017} at a distance of 8.2~kpc \citep{gravity2019}. Surrounding Sgr A$^{*}$ out to radii of $\sim$100--200 pc is a twisted torus-like structure detected at far-infrared wavelengths \citep{molinari2011}, rich in molecular gas and dust known as the Central Molecular Zone \citep[CMZ;][]{mills2017}, thought to be driven and sustained by the innermost region of the Galaxy's barred potential \citep{nayakshin2005,wardle2008,krumholz2015,tress2020}. The environment within the CMZ ($\mathrm{|}l\mathrm{|}$~$<$~0.7$^{\circ}$, $\mathrm{|}b\mathrm{|}$~$<$~0.2$^{\circ}$) is, by numerous metrics, extreme compared to that of the disk. The density \citep{,martin2004,stark2004,mills2018}, temperature \citep{guesten1985,huettemeister1993,ginsburg2016,krieger2017}, and turbulent velocity \citep{bally1987,shetty2012,kauffmann2017} of the gas in this region are all between 1--3 orders of magnitude higher than the average of the material in the disk, and the cosmic ray energy density is also 2--3 orders of magnitude higher \citep{oka2019} than that of the Galactic disk.

The region is host to several clusters of young, massive stars, including the well-known Arches \citep{cortera1996,espinoza2009} and Quintuplet \citep{nagata1990,liermann2009} clusters, as well as the stars that reside in the young, massive nuclear star cluster within the central $\sim$0.3~pc surrounding Sgr A$^{*}$ \citep{krabbe1991,maness2007,lu2009,lu2013}. The formation of these is assumed to be caused by the episodic infall of molecular clouds from the innermost orbits of the CMZ \citep{kruijssen2015,sormani2020} towards the vicinity of Sgr A$^{*}$. Some of the giant molecular cloud complexes in the GC are sites of vigorous star formation (e.g. Sagittarius B2), however this is not true for all of them, with the surprising lack of star formation in some cloud complexes being explained either due to the young ages of the clouds, or owing to the formation of higher density cores being prevented due to turbulence and strong tidal effects \citep[e.g.][]{lu2019}.

The processes whereby the infall of material, and subsequent star-formation episodes, are suppressed by radiative and mechanical ``feedback" processes remains a key theme in understanding the formation and evolution of galaxies across cosmic time. The proximity of the GC offers a unique opportunity to study these processes at high spatial \citep[and indeed temporal, e.g.][]{ponti2017} resolution. There is much observational evidence for large-scale outflows that are driving material out of the Galactic disk to higher latitudes. At scales of $\sim$100s of parsecs above or below the disk, such evidence includes but is not limited to: bipolar dust shells visible at 8.3 $\mu$m \citep{bland-hawthorn2003}; radio lobes or bubbles visible via their synchrotron emission at $\sim$GHz frequencies \citep{sofue1984,heywood2019}; the bipolar X-ray `chimneys' revealed by \emph{XMM-Newton} imaging \citep{ponti2019}; ionized thermal gas traced by radio recombination line emission \citep{law2009,alves2015,nagoshi2019}.  

The examples at radio, infrared, and X-ray wavelengths cited above are interconnected \citep{ponti2021}, and in all cases are thought to be driven by episodic, or quasi-continuous starburst or SMBH-related feedback from the Galactic center, with winds driven by high cosmic ray pressure \citep{yusef-zadeh2019}. 

On a larger than kpc scale, anomalous-velocity neutral hydrogen (H{\sc i}) clouds seen to trace a conical outflow out to Galactic latitudes of $|b|$$\sim$10$^{\circ}$ \citep{diteodoro2018} offer strong evidence that these outflows are entraining enriched material \citep{diteodoro2020}. These features bridge the gap between the Galactic center and the \emph{Fermi} bubbles \citep{su2010}, gamma-ray emitting structures with approximately coincident polarized synchrotron emission \citep{carretti2013} that extend to $|b|$$\sim$50$^{\circ}$, also thought to be caused by previous energetic outflows from the GC.

In addition to studying these larger-scale features, high resolution interferometric radio imaging has been a key method for studying the inner region of the Galactic center. Multiple radio continuum surveys of the inner $\sim$2 square degrees of the GC have been undertaken at a range of wavelengths \citep[e.g.][at 4~m, 90~cm, 20~cm and 6~cm respectively]{brogan2003,nord2004,yusef-zadeh2004,law2008}. Radio observations are not susceptible to dust obscuration, and are one of the best methods for studying the non-thermal synchrotron processes that arise due to the interaction between relativistic electrons and magnetic fields, as well as the thermal radio emission originating from ionized gas surrounding regions of star formation. One of the most notable discoveries that arose form radio observations of the GC was that of the non-thermal radio filaments \citep{yusef-zadeh1984}. These are a population of highly linear, polarized, synchrotron-emitting features that are apparently unique to the GC region.

In this article we present a new survey of the Galactic center conducted with the South African MeerKAT\footnote{Operated by the South African Radio Astronomy Observatory (SARAO).} radio telescope \citep{jonas2016} at a frequency of 1.28 GHz. The two principal data products are a total intensity mosaic covering 6.5 square degrees with an angular resolution of 4\dss, and a corresponding image of the in-band radio spectral index measurements derived from the 800~MHz of instantaneous bandwidth. In Section \ref{sec:observations} we describe the observations and data processing methods. Section \ref{sec:data_products} presents and validates the radio images. The capabilities of MeerKAT provide new insight into the morphology of many well-known radio features in the Galactic center region, as well as revealing some previously unknown structures, and offer enhanced radio views and new detections of numerous features previously imaged at other wavebands. In Section \ref{sec:discussion} we highlight some initial results from the survey, including some newly-discovered potential radio supernova remnants, high fidelity imaging of several known and new non-thermal filament complexes, and new views of the Radio Arc Bubble, Sagittarius A (Sgr A) and Sagittarius B (Sgr B) regions, the latter including the vigorously star-forming giant molecular cloud complex Sgr B2, and the former showing the inflow and outflow of material in the inner few tens of parsecs surrounding the central SMBH. The paper concludes in Section \ref{sec:conclusion}. 

\section{Observations and data processing} 
\label{sec:observations}

\subsection{MeerKAT observations}

The Galactic center region was observed during MeerKAT's commissioning phase, using the L-band (856--1712 MHz) receivers, between May and July 2018. The observations presented here consist of a 16-pointing hexagonal mosaic covering the central region, as well as four additional outlier fields. The number of dishes used for each observation was between 60 and 62 inclusive. The hexagonal mosaic is reasonably close packed, with adjacent pointings along constant Declination spaced by 0.41 degrees for a primary beam full-width at half maximum of 1.1 degrees at the band center. This was done with polarimetric studies of the region in mind, an aspect of the data that is not presented here. A summary of the pointings is given in Table \ref{tab:observations}. Note that the MeerKAT Galactic center survey also included outlier fields \citep[including those presented by][]{heywood2019} that are not included in this article. For completeness we list these remaining pointings in Appendix \ref{appendix}.

All observations were made with the MeerKAT correlator configured to deliver 4,096 spectral channels across 856--1712 MHz, recording all four polarization products. The correlator integration time per visibility point was 8 seconds for all observations except the first one for which a 4 second integration time was used.

Each block contains 10.8 hours of data (with approximately an hour of observing time lost due to the interrupted GCX15 track) for a total observing time of 215 hours. The calibration strategy during scheduling was conservative, reasoning that self-calibration of the field might prove difficult. This concern proved to be somewhat unfounded, as will be elaborated on below. The secondary calibrator 1827$-$360 was observed for 80 seconds after every 10 minute target scan. This calibrator was selected for its brightness, being 8 Jy at 1.28 GHz. After every four calibrator--target cycles, the primary calibrator PKS B1934$-$638 was observed for 10 minutes, followed by a 10 minute scan on 3C 286 to enable polarization calibration. The final on-target time per block is approximately 7.2 hours, for a total of 144 hours of data going into the final radio mosaic.

\begin{table*}
\begin{minipage}{175mm}
\centering
\caption{Summary of the MeerKAT observations used in this paper, listed in chronological order. For each observation the correlator was configured to deliver 4,096 spectral channels. The integration time per visibility point was 8 seconds, with the exception of the first block (1525645867) which had a 4 second integration time. The positions of the pointing centers are shown by the ``+" markers in Figure \ref{fig:mosaic_annotated}. Note that the observation of GCX15 was interrupted, and thus occupies two block IDs in the archive.}
\begin{tabular}{lllcccrr} \hline \hline
Date   & Block ID & N$_{\mathrm{ant}}$ & Name & \multicolumn{1}{c}{RA}              & \multicolumn{1}{c}{Dec}          & \multicolumn{1}{c}{$l$}     & \multicolumn{1}{c}{$b$}   \\ 
       &          &                    &      & \multicolumn{1}{c}{(hh:mm:ss.s)}    & \multicolumn{1}{c}{(dd:mm:ss.s)} & \multicolumn{1}{c}{(deg)}   & \multicolumn{1}{c}{(deg)} \\ \hline
2018-05-06 & 1525645867 & 62 & TARGET & 17\hhh45\mmm15\fs5 & $-$28\ddd47\dmm35\farcs3 & 0.081 & 0.142 \\
2018-06-06 & 1528308058 & 60 & GC17 & 17\hhh43\mmm35\fs9 & $-$28\ddd47\dmm35\farcs4 & 359.891 & 0.452 \\
2018-06-07 & 1528394150 & 60 & GC16 & 17\hhh44\mmm19\fs2 & $-$29\ddd06\dmm18\farcs3 & 359.708 & 0.154 \\
2018-06-08 & 1528480450 & 60 & GC23 & 17\hhh45\mmm58\fs7 & $-$29\ddd06\dmm18\farcs3 & 359.897 & $-$0.155 \\
2018-06-09 & 1528566905 & 61 & GC22 & 17\hhh45\mmm15\fs5 & $-$29\ddd25\dmm01\farcs3 & 359.549 & $-$0.183 \\
2018-06-10 & 1528652884 & 60 & GC15 & 17\hhh43\mmm35\fs9 & $-$29\ddd25\dmm01\farcs3 & 359.360 & 0.125 \\
2018-06-11 & 1528739165 & 61 & GC29 & 17\hhh46\mmm55\fs0 & $-$29\ddd25\dmm01\farcs3 & 359.737 & $-$0.492 \\
2018-06-12 & 1528824954 & 62 & GC25 & 17\hhh45\mmm58\fs7 & $-$28\ddd28\dmm52\farcs3 & 0.429 & 0.170 \\
2018-06-13 & 1528911055 & 61 & GC18 & 17\hhh44\mmm19\fs2 & $-$28\ddd28\dmm52\farcs3 & 0.239 & 0.481 \\
2018-06-14 & 1528999219 & 61 & GC31 & 17\hhh46\mmm55\fs0 & $-$28\ddd47\dmm35\farcs4 & 0.270 & $-$0.169 \\
2018-06-15 & 1529082657 & 62 & GC30 & 17\hhh47\mmm38\fs2 & $-$29\ddd06\dmm18\farcs3 & 0.085 & $-$0.465 \\
2018-06-16 & 1529168752 & 62 & GC09 & 17\hhh42\mmm39\fs6 & $-$29\ddd06\dmm18\farcs3 & 359.517 & 0.463 \\
2018-06-17 & 1529254859 & 62 & GC33 & 17\hhh46\mmm55\fs0 & $-$28\ddd10\dmm09\farcs5 & 0.803 & 0.155 \\
2018-06-18 & 1529340951 & 61 & GC32 & 17\hhh47\mmm38\fs2 & $-$28\ddd28\dmm52\farcs3 & 0.617 & $-$0.142 \\
2018-06-19 & 1529427055 & 62 & GC21 & 17\hhh45\mmm58\fs7 & $-$29\ddd43\dmm44\farcs2 & 359.364 & $-$0.480 \\
2018-06-20 & 1529521526 & 60 & GC14 & 17\hhh44\mmm19\fs2 & $-$29\ddd43\dmm44\farcs2 & 359.176 & $-$0.172 \\
2018-06-23 & 1529772166, 1529786044 & 61 & GCX15 & 17\hhh41\mmm36\fs5 & $-$30\ddd09\dmm56\farcs4 & 358.496 & 0.098 \\
2018-06-26 & 1530030351 & 60 & GCXS16 & 17\hhh40\mmm55\fs9 & $-$29\ddd29\dmm23\farcs7 & 358.992 & 0.579 \\
2018-06-29 & 1530288951 & 60 & GCX32 & 17\hhh49\mmm58\fs9 & $-$28\ddd25\dmm27\farcs9 & 0.933 & $-$0.555 \\
2018-07-01 & 1530461748 & 60 & GCX21 & 17\hhh46\mmm36\fs7 & $-$30\ddd16\dmm24\farcs3 & 358.970 & $-$0.880 \\
\hline
\end{tabular}
\label{tab:observations}
\end{minipage}
\end{table*}

\subsection{MeerKAT processing}
\label{sec:meerkat_processing}

For this project we returned to the raw archived data to ensure consistent processing, reusing none of the products from the imaging presented by \citet{heywood2019}. For each of the 20 observations, the processing steps were as follows. 

\subsubsection{Flagging and reference calibration}

The archived raw visibilities were converted to Measurement Set format using the KAT Data Access Library ({\sc katdal}\footnote{\url{https://github.com/ska-sa/katdal}}). Basic flagging commands were applied to all fields, including the bandpass edges, the Galactic neutral hydrogen line, and on baselines shorter than 600~m, the frequency ranges of known telecommunications and geolocation services. Auto-flagging procedures were then run on the uncalibrated data of the calibrator scans to remove radio frequency interference (RFI).

We derived delay (per scan), complex gain (per integration time, no frequency dependence) and bandpass (per scan, per channel) solutions from the primary calibrator PKS B1934$-$638 using the model of \citet{reynolds1994} to describe its spectrum (see also \citealt{heywood2020}). The calibration was done iteratively, and after each iteration the calibrator data were corrected, and the flagging was re-done based on the residual (model $-$ corrected) visibilities. We derived an intrinsic model for the secondary calibrator 1827$-$360  by applying the gain corrections derived from the primary, deriving complex gain corrections for the secondary in eight spectral bins, and then scaling those gains based on a corresponding set of gain corrections derived from the primary. 

The resulting corrected visibilities for the secondary were then fitted with a polynomial in frequency. This was done for each observation to account for any variability in the secondary calibrator. After the intrinsic model was fitted, additional delay and time-dependent gain corrections were derived for each scan of the secondary. Again this was done iteratively, with rounds of residual flagging of the secondary calibrator in between. Following all of these calibration steps we obtained corrections for the instrumental delays and frequency-independent gains of the telescope every 10 minutes, and bandpass corrections approximately once per hour. The target data were corrected using the solutions derived from the calibrators, and the corrected visibilities were automatically flagged. 

All calibration operations were performed using the {\sc casa} \citep{mcmullin2007} package. Calibrator flagging made use of {\sc casa}, and the target data were flagged using a custom strategy for the {\sc tricolour}\footnote{\url{https://github.com/ska-sa/tricolour/}} \citep{hugo2021} package.

\subsubsection{Imaging and self-calibration}

Following referenced calibration and flagging, the target data were split out and a blind deconvolution was performed with the sole purpose of making an initial cleaning mask for subsequent imaging. The mask was made manually using a threshold, above which no spurious features are included. We then performed a second round of imaging with the mask in place, and the resulting model was used for a round of self-calibration. Phase-only corrections were derived for every 64 seconds of data, followed by an amplitude and phase correction per scan, with no frequency dependence in the solutions in either case.

There are two main considerations when performing (self-)calibration of interferometric data. The first is the selection of appropriate solution intervals in time and frequency that strike a balance between being small enough to track the variation in those two domains of the effect(s) that is (are) being solved for, and large enough to maximise the signal to noise ratio in the solutions. The second is the completeness of the sky model. Although such models are fundamentally incomplete on some level, for complex fields it can be difficult to capture diffuse structures in the model. This is particularly true when the model is constructed using clean-based deconvolution techniques, the single-scale variant of which is notoriously poor at handling extended structures. It is now well known that incomplete sky models during calibration can lead to the suppression of unmodeled structures \citep[e.g.][]{sardarabadi2019} as well as the creation of spurious ghost sources \citep[e.g.][]{grobler2014}. To guard against this somewhat, a fairly harsh constraint was enforced that excluded spacings below 300~m from the calibration process, leveraging the longer spacings and the more readily modelled structures that they are sensitive to.

Following self-calibration the imaging proceeded iteratively, with the thresholded mask being refined after each iteration. Between 2 and 4 imaging cycles were performed, depending on the complexity of the field being processed. Following deconvolution, the multiscale model image was convolved with a 4\dss~circular Gaussian. For the residual image, a homogenization kernel was computed using {\sc pypher} \citep{boucaud16} that brings the resolution of the residuals to that of the restored model, under the assumption that the fitted restoring beam is a close approximation of the main lobe of the synthesised beam (the true point spread function). Following these two convolution operations, the model and residual images were summed.

Once the full-band, multifrequency synthesis (MFS) image was complete, we derived a final cleaning mask and then used this to image the data in 16 frequency bands. The purpose of this cube is to fit for the spectral index (Section \ref{sec:alpha_mosaic}). A hard cut (164 wavelengths) was therefore applied to the inner region of the $u,v$ plane to prevent the shortest spacings from detecting emission at the lowest end of the band that is invisible on the same physical baselines at higher frequencies, and thus artificially steepening the spectra of diffuse structures. A Gaussian taper was also applied to the gridded visibilities in each of the 16 frequency planes, to enforce a resolution for each sub-band image that was as close to 8\dss~as possible, followed by convolution of the image products. Note that 8\dss~is the target resolution here, as unlike the MFS images the sub-band images are limited by the lowest resolution in the cube. Note also that we have not combined any of the interferometer data with appropriate single dish data in order to recover the zero-spacing information.

All imaging was done using {\sc wsclean} \citep{offringa2014} with multiscale cleaning \citep{offringa2017} enabled, and a \citet{briggs95} robust value of $-$1.5, to provide high ($\sim$4\dss) angular resolution and suppress the sidelobes of the synthesised beam. Large images (10240$^{2}$ pixels with 1.1\dss~pixel size) were produced, as for every pointing there was strong emission present in the primary beam sidelobes. Frequency dependence of the sky (largely an apparent variation due to the scaling of the antenna primary beam response across the band) was captured when making the MFS images by deconvolving in eight sub-bands (as opposed to the 16 independent sub-band images that were produced for the spectral index images). The multi-scale clean component models were fitted with a fourth-order polynomial for subtraction during the major cycle. The total numbers of clean iterations across the different fields were between half a million and a million. All data processing up to this point was performed either on the Centre for High Performance Computing (CHPC\footnote{\url{https://www.chpc.ac.za/}}) Lengau cluster, or the ilifu\footnote{\url{http://www.ilifu.ac.za}} cloud computing facility. In the interests of reproducibility, and for more details on the data reduction, the scripts that deploy all the operations above on these two HPC facilities are available online\footnote{\url{https://github.com/IanHeywood/oxkat} v0.1} \citep{oxkat2020}. 

We applied primary beam correction to both the full-band MFS images and the 16 sub-band spectral cubes. In the case of the former, the nominal band center frequency (1284~MHz) was used. The primary beam correction was a simple image-plane operation, whereby the image is divided by a model of the primary beam at the appropriate frequency. In this case we used the {\sc eidos} package \citep{asad2021} to evaluate the Stokes I beam at the relevant frequencies, and then azimuthally-averaged this product to remove the non-circular asymmetries in the main lobe of the MeerKAT primary beam. The images were blanked beyond the point where the beam gain nominally drops below 0.5.

\subsection{VLA observations and processing}
\label{sec:vlaobs}

We also made use of observations from the Karl G.~Jansky Very Large Array (VLA) in order to validate the MeerKAT data products. These were two observations\footnote{Project codes: 15A-286 and 15A-310} using L-band (1--2 GHz) with the array in the most extended A-configuration. The observations targeted the Sgr A (J2000 17\hhh45\mmm40\fs0 $-$29\ddd00\dmm28\farcs0) and Sgr C (J2000 17\hhh44\mmm35\fs0 $-$29\ddd29\dmm00\farcs0) regions, with 5.56 and 2.5 hours of on-source time respectively.

The initial flagging and reference calibration was performed using the VLA {\sc casa} pipeline\footnote{\url{https://science.nrao.edu/facilities/vla/data-processing/pipeline}}. Following this we performed some quick but coarse flagging by identifying scan / spectral-window pairs that had anomalously high visibility amplitudes and flagging them outright. The processing that followed was akin to the MeerKAT processing. The data were deconvolved blindly in order to construct a cleaning mask, and then re-imaged with the {\sc wsclean} multiscale clean algorithm. A single round of phase-only self-calibration was then performed, and the data were re-imaged. The angular resolution of the resulting images is $\sim$1\dss. Further details on the use of the VLA data are given in Section \ref{sec:astrometry}.

\section{Data products}
\label{sec:data_products}

\begin{figure*}[p!]
\centering
\includegraphics[width=\textwidth]{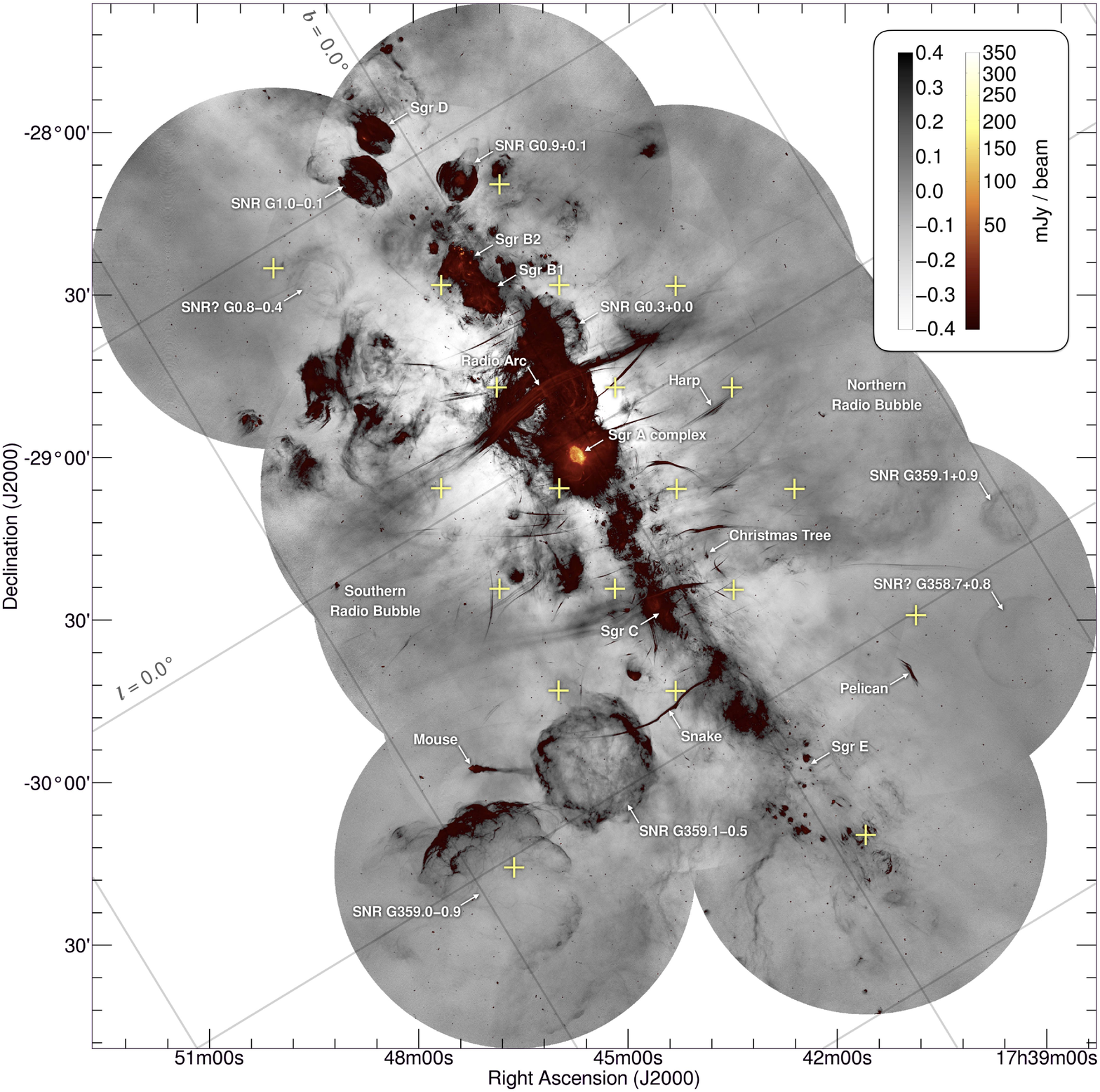}
\caption{The full MeerKAT total intensity mosaic, covering 6.5 square degrees of the Galactic center region at an angular resolution of 4\dss. This is a linear mosaic formed from the 20 pointings described in Section \ref{sec:observations}, the centers of which are shown in the figure as ``+" markers. The image has dual color schemes, with a linear greyscale covering the faint end, and the heat map covering the bright end with a square-root stretch function. Pixel scales for the two colormaps are inset. Several radio features in the region are highlighted, including the Mouse \citep{yusef-zadeh1987}, the Snake \citep{gray1991} and the Pelican \citep{lang1999}, the Harp and the Christmas Tree \citep{thomas2020}, and numerous supernova remnants as cataloged by \citet{green2019}. The Radio Arc \citep{yusef-zadeh1984,pare2019} is highlighted, now known to be coincident with part of the eastern boundary of the 430 pc bipolar radio bubbles that span the Galactic center \citep{heywood2019}. These are also annotated, although their full extent is not covered by the image above. An additional grid showing Galactic coordinates can be found on the figure. The lines corresponding to Galactic coordinates $l$~=~0$^{\circ}$ and $b$~=~0$^{\circ}$ are marked, and the associated grid spacing is 1$^{\circ}$. We discuss some features in this image in further detail in Section \ref{sec:discussion}.\vspace{5mm}
\label{fig:mosaic_annotated}}
\end{figure*}

\subsection{Total intensity mosaic}

A linear mosaic was formed from the 20 primary-beam-corrected MFS images using the {\sc montage}\footnote{\url{http://montage.ipac.caltech.edu}} software package. A rendering of the full mosaic is presented in Figure \ref{fig:mosaic_annotated}. To capture the high dynamic range on the mosaic, it is shown with a dual color scheme, whereby the faint end is covered by a linear greyscale, and the bright end is covered by a heat map with a square-root pixel stretch. The overall pixel scale is presented on the figure. The standard practice of using variance weighting was adhered to, with the square of the primary beam attenuation pattern being used as a weighting function. In practice the images are deconvolution limited, and not limited by a thermal noise background, and some minor discontinuities can be seen at the pointing boundaries. The total area covered by the mosaic is 6.5 square degrees. The 20 pointing centers are shown by the ``+" markers, and the figure is annotated to highlight some known radio features in the Galactic center region (refer to the figure caption for further details). We discuss some of the features in this image further in Section \ref{sec:discussion}, in which we present some subsets of the mosaic with quantitative color scales.

\subsection{Astrometric corrections}
\label{sec:astrometry}

A small number of bugs arose in the processing of early commissioning data from MeerKAT, that gave rise to systematic inaccuracies in the measured positions of radio sources \citep[see also][]{mauch2020,knowles2021}. Briefly, the two potential issues were: (i) the timestamps in early data from the array were erroneously offset by a single correlator-beamformer interval (2 s), leading to $u,v,w$ coordinate errors, and in turn an apparent rotation of the field about the phase center; (ii) inaccurate positions of some calibrator sources were employed, breaking the ``central point source'' assumption that standard calibration routines make, and leading to a corresponding offset in the target data when the complex gain corrections derived from the calibrators were applied to the science target.

Undertaking a consistent reprocessing of the data allowed issue (i) to be corrected for during the conversion of the raw visibilities to Measurement Set format using a fixed version of {\sc katdal}. The telescope was commanded to observe the position J2000 18\hhh30\mmm58\fs8 $-$36\ddd02\dmm30\farcs1 for scans of the secondary calibrator 1827$-$360. This source does not have an entry in the International Celestial Reference Frame catalog \citep[v3,][]{charlot2020}, however the position used is offset from the position in the Australia Telescope Compact Array (ATCA) calibrator database, by 1\farcs01 in RA, and 0\farcs03 in Dec. To investigate any astrometric issues caused by this (or other unforeseen issues), we make use of the VLA observations targeting Sgr A and Sgr C that we introduced in Section \ref{sec:vlaobs}. The {\sc pybdsf} source finding software \citep{mohan15} was run on the two VLA images with its default settings. {\sc pybdsf} works by using a sliding window to estimate the background noise level in the map as a function of position ($\sigma_{\mathrm{local}}$), and then finding peak pixel values that exceed some multiple of this background (in this case 5$\sigma_{\mathrm{local}}$). A flood-fill algorithm then determines regions of contiguous emission down to some secondary threshold (3$\sigma_{\mathrm{local}}$ by default) to identify islands. These islands are then decomposed into a list of point and Gaussian components, which are exported as a catalog. Since the VLA images are synthesised from observations with the most extended A-configuration which lacks significant numbers of short baselines, several hundred compact features in the image can be reliably identified by automatic source-finders as there is no strong, large-scale emission to confuse the background noise estimation.

The VLA component list returned by {\sc pybdsf} was filtered to include only islands of emission that were fitted by a single point or Gaussian component. We created cut-out images spanning 72\dss~at the positions of these sources, extracted from the MeerKAT mosaic. These were visually examined to create a list of MeerKAT sources that could also be reasonably fitted by a single point or Gaussian component, with the additional criterion that the compact feature was embedded in a uniform background. With these criteria met, we obtained a list of 144 compact MeerKAT sources with VLA-detected counterparts that can be used to test the astrometric accuracy.

\begin{figure}[h]
\centering
\includegraphics[width= \columnwidth]{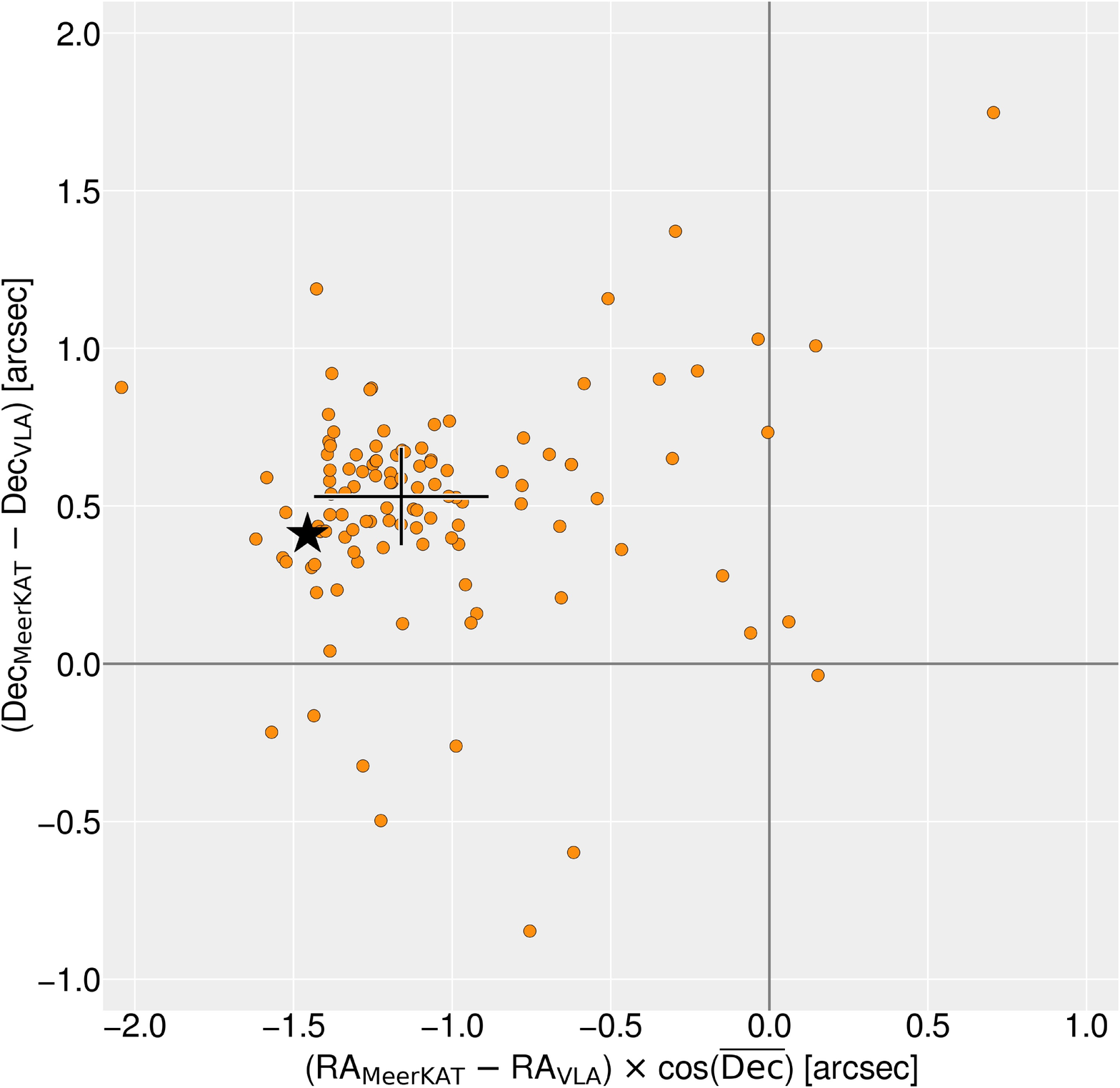}
\caption{Offsets in RA and Dec for 144 compact radio sources that are detected in both the MeerKAT mosaic, and a pair of VLA L-band observations with 1\dss~angular resolution targeting the Sgr A and Sgr C regions. The median offsets in RA and Dec are $-$1\farcs16 and 0\farcs53, with interquartile ranges of 0\farcs27 and 0\farcs15 respectively, as shown by the position and extent of the large `+' symbol. The star symbol shows the offset between the Very Long Baseline Array position of Sgr A$^{*}$ at 24~GHz \citep{petrov2011}, and the MeerKAT measurement. The median measured offsets are used to bring the MeerKAT astrometry in line with that of the VLA. See Section \ref{sec:astrometry} for further details.}
\label{fig:astrometry}
\end{figure}

The distribution of the measured offsets is shown in Figure \ref{fig:astrometry}, in addition to the star symbol which shows the Very Long Baseline Array position of the compact radio source Sgr A$^{*}$ \citep{petrov2011}. The cross marks the median of the distribution ($\tilde{\Delta}_{\mathrm{RA}}$ = $-$1\farcs16, $\tilde{\Delta}_{\mathrm{Dec}}$ = 0\farcs53) and the extent of that marker shows the interquartile ranges which are 0\farcs27~and 0\farcs15~in RA and Dec respectively. The offset in RA is consistent within the errors to the offset between the MeerKAT pointing position for the secondary calibrator and the position in the ATCA calibrator database. The cause of the offset in Dec between the VLA and MeerKAT detections remains unknown. We bring the MeerKAT astrometry in line with the higher angular resolution VLA data by applying these offsets to the headers of the mosaic products, however users should be aware that astrometric errors on the sub-pixel level may remain. A catalog of previously unidentified compact radio sources is under construction, and will be presented in a companion paper (Rammala et al., \emph{in prep.}), thus the compact source catalogs used for the astrometric corrections are not presented here. 

\subsection{Spectral index mosaic}
\label{sec:alpha_mosaic}

\begin{figure}[h]
\centering
\includegraphics[width= \columnwidth]{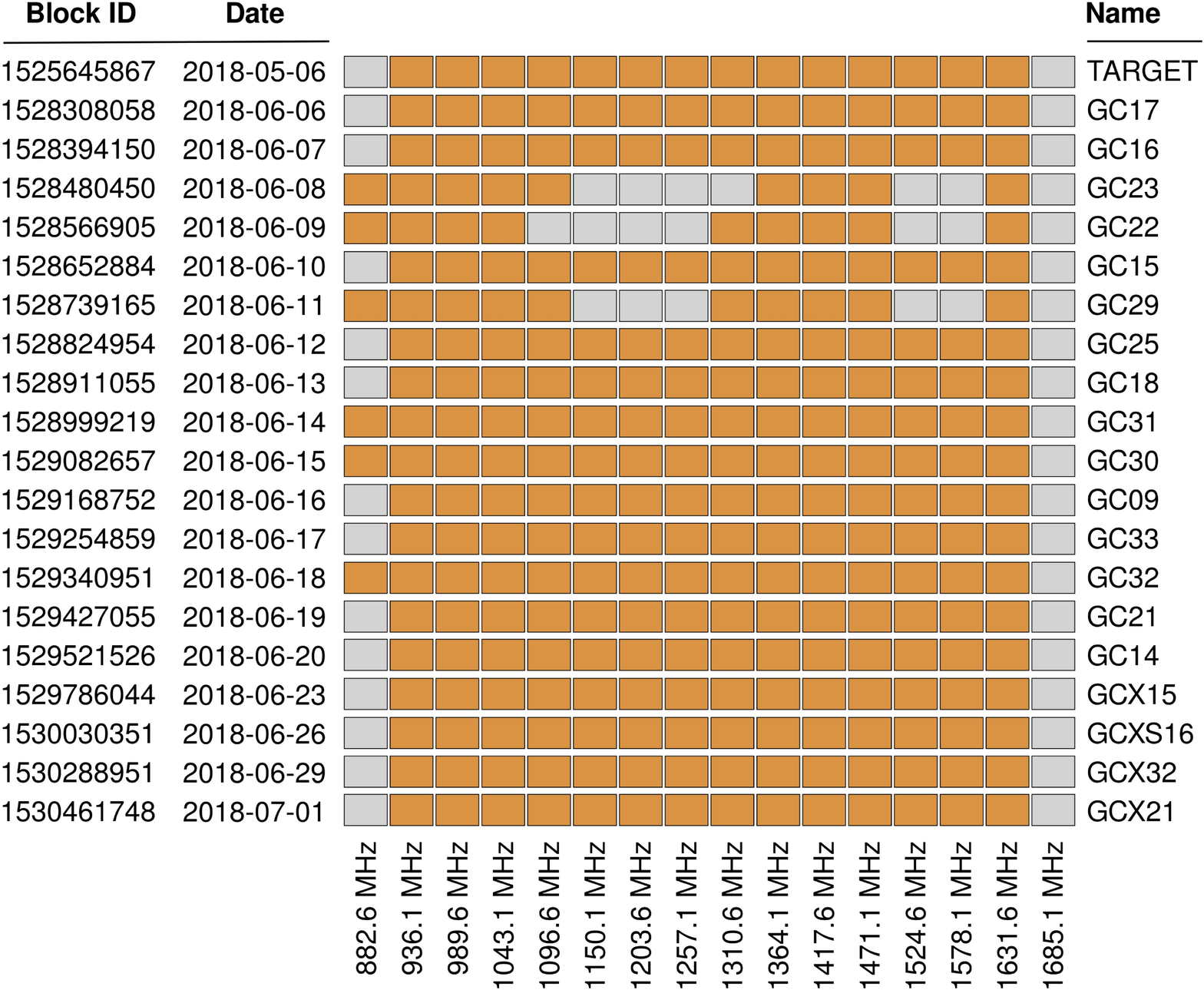}
\caption{A chart showing which of the 16 sub-bands were missing (the gray squares) from each of the 20 pointings when forming the sub-band mosaics for producing the spectral index mosaic. The central frequency of each 53.5\,MHz wide sub-band is noted on the horizontal axis. Refer to Section \ref{sec:alpha_mosaic} for further details.}
\label{fig:subbands}
\end{figure}

We visually examined the 16~$\times$~53.5~MHz sub-band images for each of the 20 pointings, and removed images where the sensitivity or image fidelity was compromised due to RFI losses. This reduced the number of sub-band images from 320 to 269. In all cases the highest frequency sub-band ($\nu_{\mathrm{center}}$~=~1685.1~MHz) was lost, and in all but six pointings the lowest sub-band ($\nu_{\mathrm{center}}$~=~882.6~MHz). RFI compromised the image quality of numerous sub-bands, mainly through the loss of shorter spacings, however it resulted in the total loss of only 17 additional sub-bands, spread across three pointings. A visual summary of the frequency coverage on a per-pointing basis is provided in Figure \ref{fig:subbands}. For each sub-band the primary beam corrected images were mosaicked using the same method as for the full-band total intensity mosaic. The resulting mosaics were then assembled into a cube. The spectrum (flux density against frequency) for each line of sight through the cube was extracted, and a linear fit to the slope of this spectrum in log space was performed. The slope and the associated error are the spectral index (and associated error), and these two values are written to FITS images. Note that we adopt the convention that the flux density $S_{\nu}$ is related to the frequency $\nu$ via the spectral index $\alpha$ according to $S_{\nu}~\propto~\nu^{\alpha}$. Pixel values were blanked in the Stokes I cube below 10 $\mu$Jy~beam$^{-1}$ to avoid fitting negative values. For a spectral fit to be considered for a particular sight line, fewer than half of the 16 channels must be blanked (either due to the thresholding, the primary beam cut, or RFI losses). We make no attempt to fit for spectal curvature, as in-band measurements from fractional bandwidths such as those used here are likely to be unreliable for all but the brightest components \citep[e.g.][]{heywood2016}.

This process will still fit regions of the cube that result in unphysical spectra, generally due to regions containing persistent, low level sidelobe emission from incomplete deconvolution of large angular scale emission in the Galactic plane. Any residual sidelobe features associated with incomplete deconvolution will also exhibit spectral behavior that is smooth in frequency rather than being noise-dominated, and so the error in the fit cannot be used as a reliable discriminator for certain regions. It is not possible to enforce a set of masking criteria that are satisfactory for the full mosaic due to the large dynamic range. Therefore, region-specific masking using combinations of spectral index error and flux density cuts are advised for examining the spectral index of targets of interest. Some spectral index images are presented with such masking methods applied in Section \ref{sec:discussion}. Note that as mentioned in Section \ref{sec:meerkat_processing}, the angular resolution of the cube and therefore the spectral index image is 8\dss, i.e.~a factor of 2 lower than the total intensity mosaic. This is an unavoidable limitation of one being produced using sub-band imaging and the other being produced using MFS imaging.

\section{Initial science results}
\label{sec:discussion}

The largest coherent structure to be seen in the MeerKAT mosaic (excluding the Galactic plane itself) is the 430 pc bipolar radio bubbles, the northern and southern halves of which are denoted in Figure \ref{fig:mosaic_annotated}, and are presented in detail by \citet{heywood2019}. In this section we focus on structures that are smaller in scale but nevertheless heavily-resolved in the MeerKAT data. We present a broad overview of a few regions of the mosaic that demonstrate the improved view of some well-known regions that the MeerKAT data affords, as well as introducing some new discoveries, including potential radio supernova remnants and non-thermal filament complexes. We leave discussion of the third scale category, that is compact radio features that range from point-like to a few resolution elements across, to a companion paper (Rammala et al., \emph{in prep.}), which will present a catalog of such objects.

\subsection{Supernova remnants}

\subsubsection{SNR G0.9+0.1: evidence of polar outflows from the pulsar wind nebula}

\begin{figure}[h]
\centering
\includegraphics[width= \columnwidth]{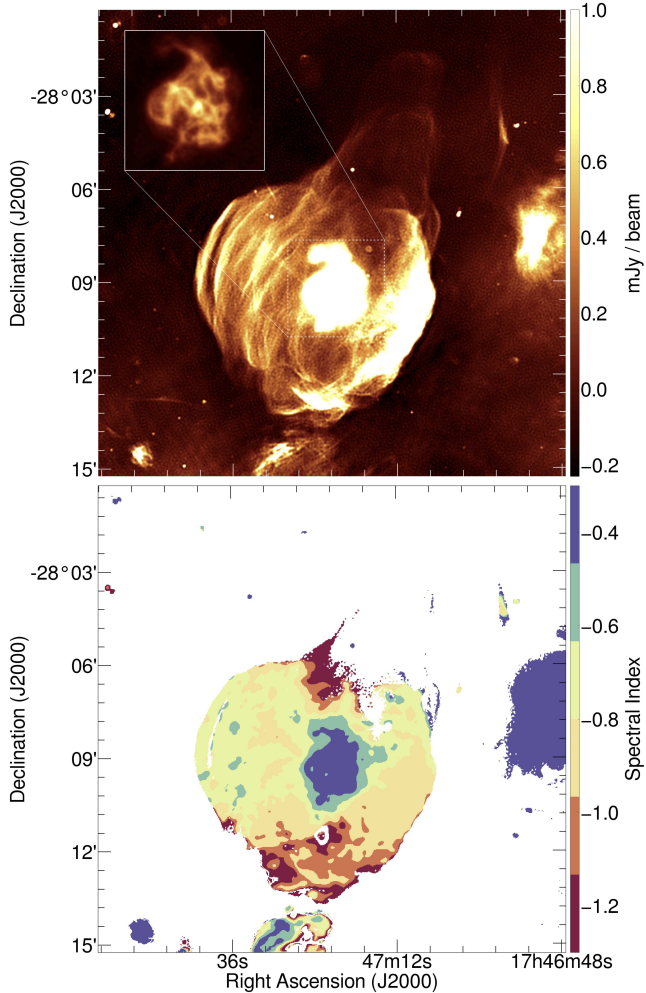}
\caption{The upper panel shows the total intensity image of SNR G0.9+0.1 with a resolution of 4\dss, and the lower panel shows the in-band spectral index measurements over the same region. The latter has been masked where the Stokes I brightness drops below 100 $\mu$Jy beam$^{-1}$. A northern ``ear" is revealed in the supernova shell, likely due to activity associated with the central pulsar wind nebula (PWN). A desaturated view of the PWN is shown inset in the top panel, revealing it to be a tangled, filamentary structure. The spectral index map highlights the flatter spectrum PWN embedded in the synchrotron emission of the SNR, and also shows a spectral index gradient consistent with the elongated axis of the SNR.}
\label{fig:snrG09}
\end{figure}

Figure \ref{fig:snrG09} shows the total intensity image of SNR G0.9+0.1 (upper panel) and the corresponding in-band spectral index measurements from the MeerKAT data (lower panel). The feature in the center of the shell was identified as a pulsar wind nebula (PWN) by \citet{gaensler2001} using X-ray observations, and a pulsar was subsequently discovered within by \citet{camilo2009}. A de-saturated view of the PWN is provided in the inset panel of Figure \ref{fig:snrG09}. Previous radio imaging by \citet{dubner2008} reported a disk and jet structure consistent with the X-ray morphology. The MeerKAT imaging reveals a complex, tangled filamentary structure, somewhat resembling the \emph{Chandra} X-ray image of the Crab nebula \citep{weisskopf2000}.

A protrusion in the shell boundary is clearly visible to the northwest of the image, which must be driven by the central PWN, and is aligned with the putative jet axis described by \citet{dubner2008}. This distortion of the shell could be due to the jet itself, excavating channels along its axis and breaking out in the north, analogous to the shell in the W50 supernova remnant that has been distorted by the jets of its central progenitor, the X-ray binary SS 433 \citep{dubner1998}. Alternatively an isotropic wind from the PWN could be collimated preferentially in the north-south direction if the original supernova explosion produced an expanding shell with bipolar density enhancements in the east-west direction (or expanded into such an environment). Such a configuration can be clearly seen in the single-dish radio observations of SN 1006 \citep{dyer2009}. The enhanced brightness, particularly on the western edge, indicates that this is a possibility. The nested, edge-brightened features in the northern protrusion are also suggestive of episodic activity from the PWN.

This activity is also reflected in the in-band spectral index map of this source, shown in the lower panel of Figure \ref{fig:snrG09} (refer to the caption for details). The PWN complex is seen as the central relatively flat-spectrum feature inside the synchrotron emitting shell, with steeper spectrum emission seen aligned with the elongation axis.

\begin{figure*}[ht!]
\centering
\includegraphics[width= 0.95 \textwidth]{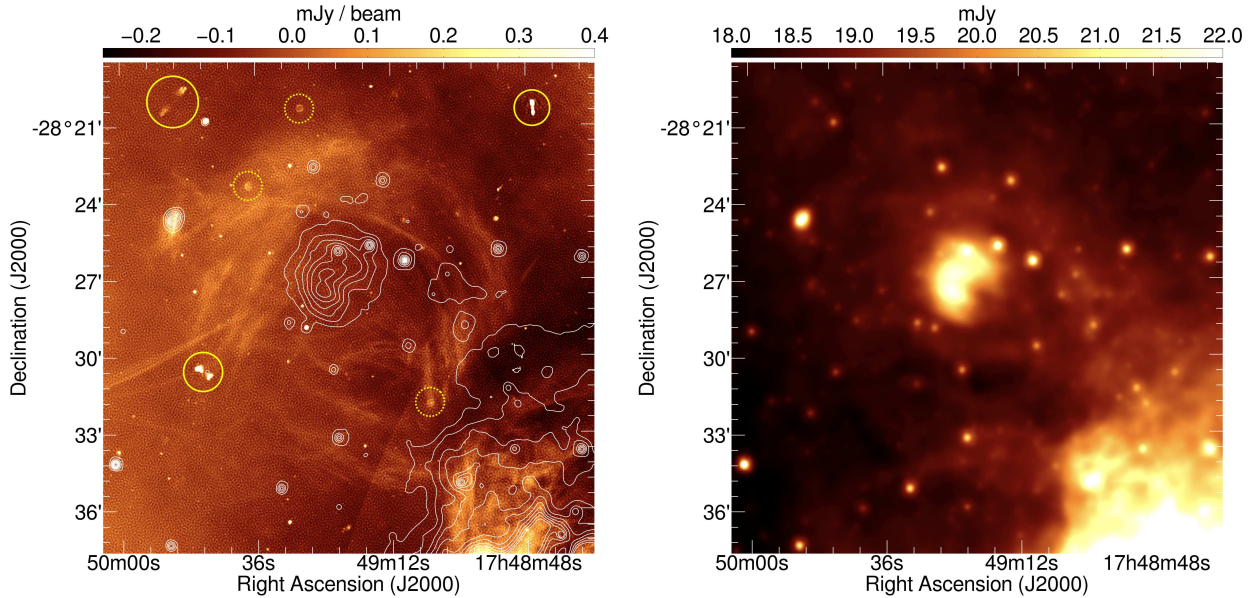}
\caption{The left hand panel shows the low surface brightness, shell-like 1.28 GHz radio structure at G0.8$-$0.4, possibly a new radio supernova remnant, or a shell driven by thermal winds from an H\,\textsc{ii} region. The contours trace the Band 4 (22 $\mu$m) WISE emission, the image of which covering the same region is shown in the right hand panel. Contour levels start at 19.15 mJy and increase linearly with a spacing of 0.5 mJy. 
Three sources that are likely background double-lobed radio galaxies are also shown in the left hand panel, highlighted by the solid circles. Three of the numerous low angular diameter shells that are visible in the MeerKAT mosaic can also be seen, highlighted by the dashed circles. The image has an angular resolution of 4\dss.}
\label{fig:G08-04}
\end{figure*}

\subsubsection{G0.8$-$0.4: a new radio SNR?}

Figure \ref{fig:G08-04} shows a low surface brightness, filamentary shell-like structure that is a possible new SNR, or perhaps a thermal wind-driven H\,\textsc{ii} bubble. The diameter of the shell is approximately 12$'$. No counterparts at other wavebands were found in any of the large-area surveys made available via the Aladin Sky Atlas \citep{bonnarel2000,boch2014}, however a source central to the radio shell is seen in the Wide-field Infrared Survey Explorer \citep[WISE;][]{wright2010} Band 4 image at 22 $\mu$m, but in no other bands. The WISE image shows a shell or partial shell-like source, with brightness enhancements on the eastern edge. The source appears to be offset from the center of the filamentary radio shell. It is cataloged (as G000.834-00.457) in the list of H\,\textsc{ii} regions identified in IRAS imaging by \citet{anderson2014}, and in the list of bubbles (as G000836-004493) identified in GLIMPSE/MIPSGAL data via a citizen science project \citep{simpson2012}. No properties other than position and morphological parameters are given in either case. No reliable in-band spectral index measurements were achievable for the filamentary shell, due to its low surface brightness. The standard deviation of the pixels in the radio image over the area of the IRAS detection are 26 $\mu$Jy beam$^{-1}$. If this central infrared source is an H\,\textsc{ii} region then the lack of corresponding radio emission in our deep imaging is somewhat puzzling. The integrated 22 $\mu$m flux of $\sim$350 Jy would mark this source as an extreme outlier in terms of its radio to infrared brightness ratio \citep{makai2017}, although there is observational evidence for dust emission interior to (and offset from) the ionized shell in spatially-resolved H\,\textsc{ii} regions \citep{hankins2019}. Supernova remnants have also been seen to exhibit far-infrared emission that is off-center from their radio shells, thought to be due to the production of dust \citep{lau2015,delooze2017}.

The three double-lobed radio sources in the field (highlighted by solid circles) have decidedly synchtrotron-like spectral indices, and are likely background active galactic nuclei (AGN). This field also shows some of the numerous low angular diameter shells (a few resolution elements across, three of which are highlighted by the dashed circles). These could be ionization regions around massive stars, or alternatively planetary nebulae (PNe). Assuming a typical PN diameter of 0.3~pc \citep{frew2016}, a 15\dss~angular diameter shell would be at a distance of 4.3~kpc. PNe that are close to the Galactic center are thus not likely to be spatially resolved in this mosaic. The deficiency that arises from mosaicking individual images that are limited by bright extended structures with slightly differing levels of deconvolution, rather than thermal noise limited, is also evident in Figure \ref{fig:G08-04}, manifesting as low-level discontinuities that trace the circular primary beam cut.

\subsubsection{G358.7+0.8: discovery of a remarkably spherical radio nebula}
\label{sec:bubble}

\begin{figure}[h]
\centering
\includegraphics[width= \columnwidth]{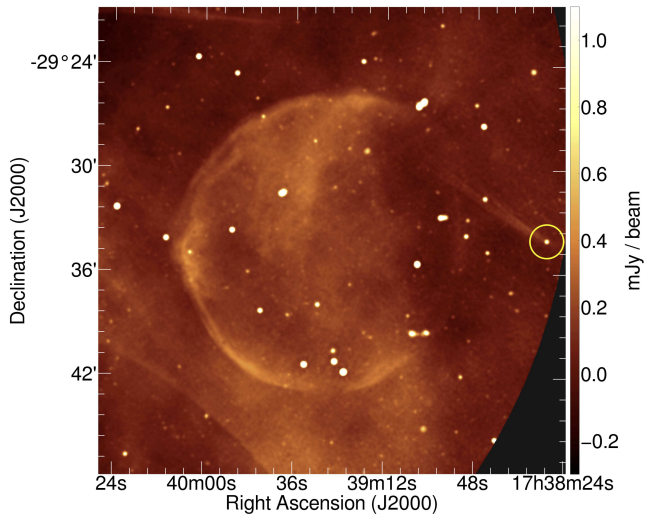}
\caption{G358.7+0.8 is a newly-discovered low surface brightness, almost spherical radio nebula. We found no multiwavelength counterparts to this object. The image above shows the MeerKAT image convolved to 11\dss~to highlight the structure. The eastern rim of the shell has a spectral index estimate that is consistent with synchrotron emission, suggesting it is a supernova remnant with rare spherical morphology. The tailed source discussed in Section \ref{sec:bubble} is highlighted by the circle.}
\label{fig:bubble}
\end{figure}

The northwestern corner of the MeerKAT mosaic features a low surface brightness radio nebula (G358.7+0.8), notable for its almost-spherical morphology. The structure is shown in Figure \ref{fig:bubble}, in which the MeerKAT image has been convolved to an angular resolution of 11\dss. We were unable to locate a counterpart to this source in multiwavelength imaging of the region \citep{bonnarel2000,boch2014}, suggesting it is a new discovery.

The eastern rim of the shell is bright enough to estimate the in-band spectral index, which is consistent with synchrotron emission, with a mean value of $\alpha$~=~$-$0.6, computed over an area of the eastern rim covering $\sim$200 restoring beams. This suggests that G358.7+0.8 may be a supernova remnant, although its position at the very edge of the mosaic means that the spectral baseline for the in-band spectral index estimate will be short. Supernova remnants with such spherical morphologies are rare however, with SN1066 likely being the best known example that is well-studied in the radio \citep{dyer2009}. 

It is unlikely that G358.7+0.8 is a planetary nebula (PN), although its radio morphology bears a close resemblence to the optical image of the spherical PN Abell 39 \citep{jacoby2001}. However the radio emission from PNe is, in essentially all cases, thought to originate from thermal free-free processes, and thus if the spectral index estimate is accurate (and holds for the entirety of the shell), it is at odds with this property. Also, the angular diameter of 17\dmm~would place the putative PN at a distance of only 61 pc, again assuming a typical nebula diameter of 0.3 pc \citep{frew2016}. This would make it by far the closest known PN, in which case the lack of a counterpart in large-area surveys at other wavelengths would be difficult to explain.

We note also the presence of a compact source with a ``Mouse-like" tail to the northwest of the shell. Given the position of this relative to the shell (and the undisturbed morphology of the shell) it is likely a chance alignment. The compact source appears to have inverted spectrum emission ($\alpha$~$\sim$~0.6). The tail in this case may be a non-thermal filament, arising as a consequence of cosmic-ray driven nuclear winds interacting with a mass-losing stellar source \citep[][see also Section \ref{sec:filaments}]{yusef-zadeh2019}.

\subsubsection{The MeerKAT view of SNR candidates G0.33+0.04 and G359.172+0.264}

\begin{figure}[h]
\centering
\includegraphics[width= \columnwidth]{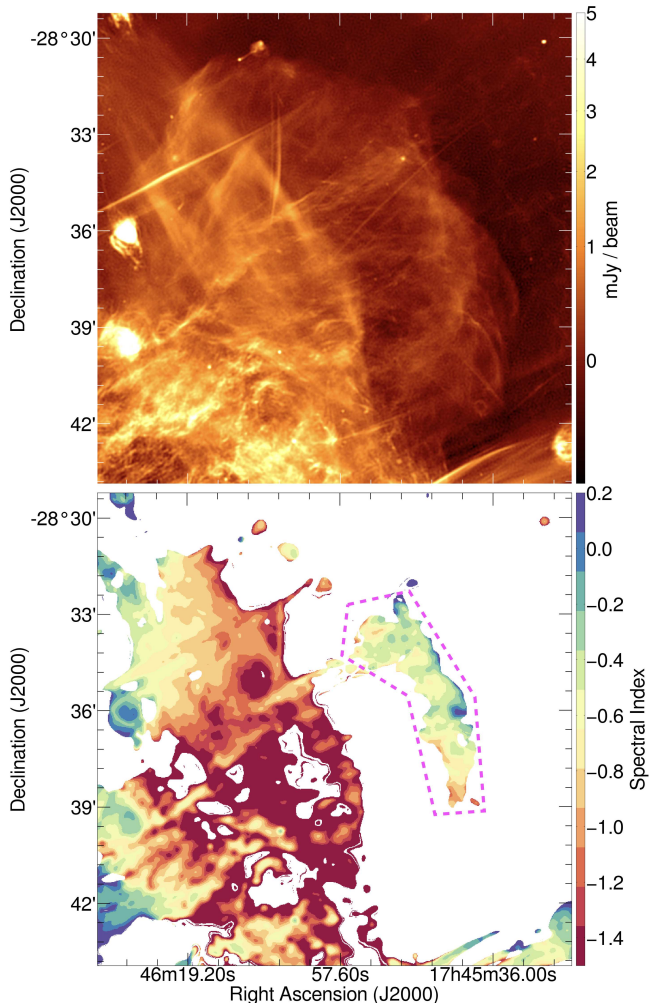}
\caption{The 1.28 GHz MeerKAT image of the SNR candidate G0.33+0.04 is shown in the upper panel, using a square root pixel stretch. A complete shell is revealed with prolate morphology, the major axis of which is approximately aligned with the Galactic plane. The in-band spectral index map is presented in the lower panel. The brighter western rim of the shell has an average spectral index (computed within the magenta box) of $-$0.5, close to the value of $-$0.56 reported by \citet{kassim1996} using multifrequency radio observations.}
\label{fig:snrG0p33}
\end{figure}

Figure \ref{fig:snrG0p33} shows the 1.28 GHz MeerKAT image of the SNR candidate G0.33+0.04, the position of which is also noted on Figure \ref{fig:mosaic_annotated}. It has been previously studied in detail by \citet{kassim1996}, who interpret the source as an SNR based on their multifrequency radio spectrum (spanning 80~MHz to 15~GHz) and its shell-like morphology. Those authors note G0.33+0.04 as of potential interest due to its apparent proximity to Sgr A$^{*}$, second only to Sgr A East in terms of known SNRs (or SNR candidates, see Section \ref{sec:30pc}).

The MeerKAT image shows a full prolate shell with a projected angular size of 0.19\ddd~$\times$~0.13\ddd~(27.5~pc~$\times$~18.8~pc), at a position angle of 35\ddd. The emission is significantly confused by complex emission from the plane in its southeastern half, and is crossed by numerous non-thermal filaments. We can recover in-band spectral index information for the western rim of the shell, as shown in the lower panel of Figure \ref{fig:snrG0p33}. The mean spectral index measured within the polygon is $-$0.5, close to the value of $-$0.56 reported by \citet{kassim1996} in their multifrequency spectrum above the turnover at $\sim$150 MHz.

We also examine the SNR candidate G359.172+0.264 reported by \citet{dakora2021} from the Global View of Star Formation in the Milky Way survey \citep[GLOSTAR;][]{brunthaler2019,medina2019}. The candidate is detected in the VLA D-configuration data from that survey, where it is morphologically classified as a filled shell with an integrated 5.8~GHz flux density of 17.5~$\pm$~4.4 mJy. Correcting this to 1.28 GHz assuming $\alpha$~=~$-$0.5 gives an integrated flux density of 37 mJy. Despite the source being situated in a localized negative bowl with a mean depth of $-$0.15 mJy beam$^{-1}$, a source with such a C-band flux density and typical SNR characteristics should be readily detectable by MeerKAT. Only a hint of a partial shell (with a brightness enhancement to the south) is seen in the data, the peak of which is at a 1.5$\sigma$ level. The unusual spectrum of this source could possibly be investigated by follow-up observations using MeerKAT's S-band receivers \citep[1.75--3.5 GHz;][]{barr2018}.

\subsection{Non-thermal filaments}
\label{sec:filaments}

Several complexes of magnetized filamentary emission have been identified over the last few decades in the inner $\sim$2 square degrees of the Galactic center. Their intrinsic polarization and spectral index show that they are synchrotron sources with magnetic fields directed along the length of the filaments \citep[e.g.][]{law2008,pare2019}. There are many theories as to the origin of the filaments \citep[e.g.][]{shore1999,boldyrev2006,linden2011,barkov2019,yusef-zadeh2019} that aim to understand both the origin of the magnetic field as well as the source of relativistic particles that drive the synchrotron emission, but there is no single, conclusive explanation as to their origin.

\begin{figure*}[p!]
\centering
\includegraphics[width=0.8 \textwidth]{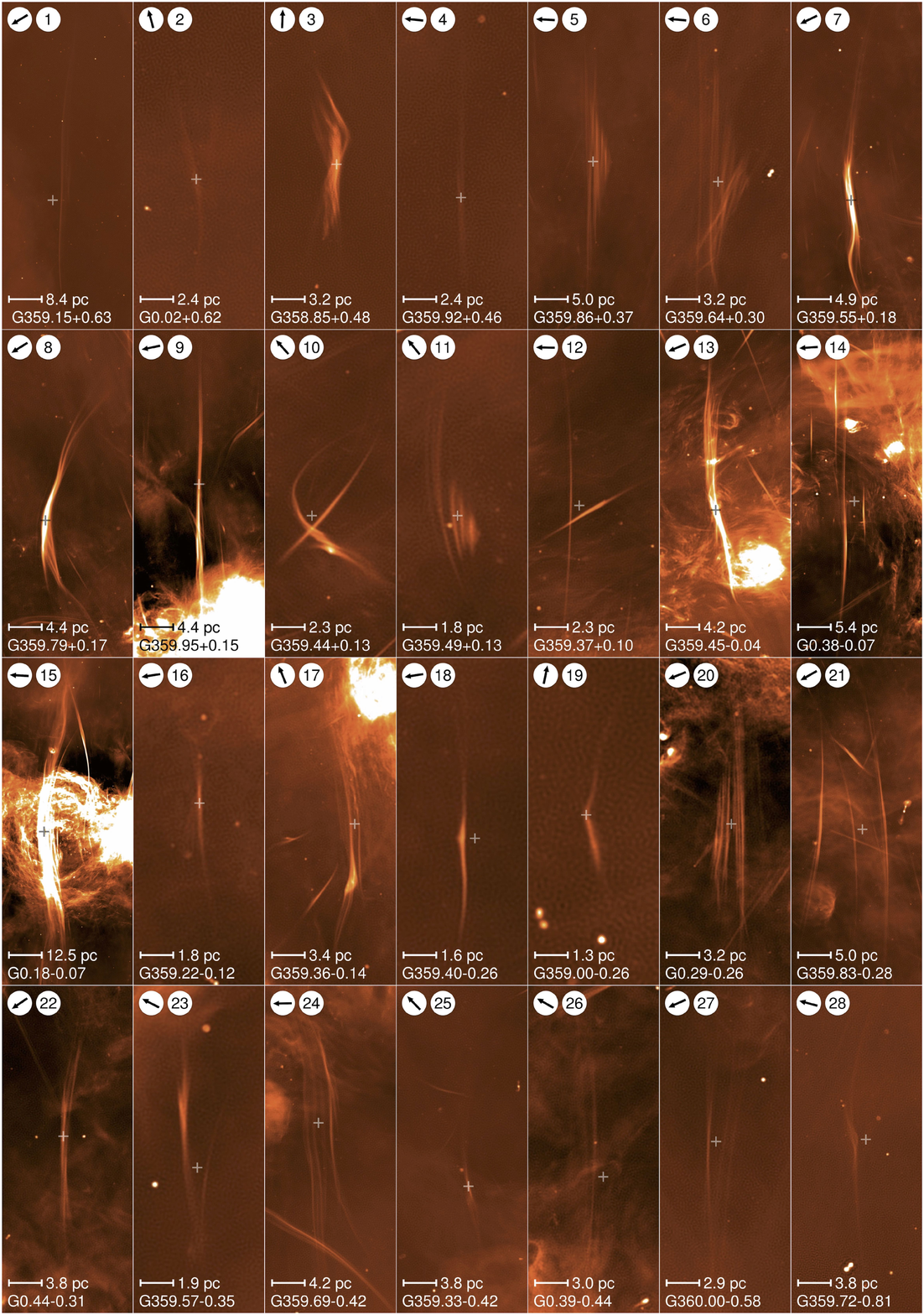}
\caption{Some examples of non-thermal radio filaments from the 4\dss\,resolution MeerKAT mosaic. For economy of space, the images have been rotated such that the principal axis of the filaments run vertically. The arrow in the top left of each panel shows the direction of increasing $l$, i.e.~it points eastward along the Galactic plane, showing its orientation relative to the filaments. The Galactic coordinates of the ``+" marker is given in each panel. This is only for finding purposes, having been moved from the center of the panel in some cases in order to not obscure interesting features, and does not represent the position of any filament, or filament complex. The brightness scale is linear, and common to all panels. The bar on each panel shows the linear scale in pc, assuming a distance to the Galactic center of 8.2~kpc \citep{gravity2019}, and that all filaments lie at this distance. Refer to Section \ref{sec:filaments} for futher details.
\label{fig:filaments}}
\end{figure*}

Figure \ref{fig:filaments} shows 28 filaments and filament complexes as seen by MeerKAT. While far from exhaustive, these examples are representative, and show the varying morphological properties within the general population. Extremely linear morphologies are the defining characteristic, however common features can be seen between separate groups of filaments. For example, filaments can appear to be isolated (panels 4, 16, 18, 19) or occur as complexes of multiple, parallel filaments (panels 5, 6, 11, 20, 24). Panels 5 and 11 respectively show ``the Harp" and ``the Christmas Tree", so-called by \citet{thomas2020}. These authors explain this morphological configuration in terms of a massive star or pulsar that is ejecting cosmic rays as it moves along Galactic magnetic field lines, which are then sequentially emitting synchrotron emission. Panel 20 shows a similar filamentary morphology, but also contains an intriguing, possibly associated, tailed (``Mouse-like") compact object ($\alpha$~=~$-$0.3) that appears to be moving away from the filaments, similar to the source seen in Figure \ref{fig:bubble}.

Curvature of various levels is seen, sometimes as undulations along the filament(s) (panels 3, 7, 8), sometimes significantly more pronounced (panel 10). Several filaments exhibit a kink along their length (panels 17, 18, 19) which is typically accompanied by a pronounced enhancement in the radio brightness. Another example of this can be seen in the well-known ``Snake" filament, that is visible and labelled in Figure \ref{fig:mosaic_annotated}. Compact radio sources often appear to be embedded in the filaments (panels 10, 11, 13, 26), however it is not known whether this is just due to chance alignment.

MeerKAT also reveals for the first time numerous low surface brightness filaments (e.g. panels 2, 4, 27, 28), including some that are situated at projected distances of $>$200 pc away from the GC. In addition to the Snake, the brightest filament complexes are the Radio Arc (panel 15) and the filaments adjacent to Sgr C (panel 13). Both of these are seen to be significantly longer than previously thought. Filaments associated with the Radio Arc complex can be seen to span $\sim$200 pc in Figure \ref{fig:mosaic_annotated} before fading below detectability, and both the Radio Arc and Sgr C filament complexes are now known to mark the boundaries of the 430 pc bipolar radio bubbles \citep{heywood2019}.

\begin{figure*}[h]
\centering
\includegraphics[width= 0.93 \textwidth]{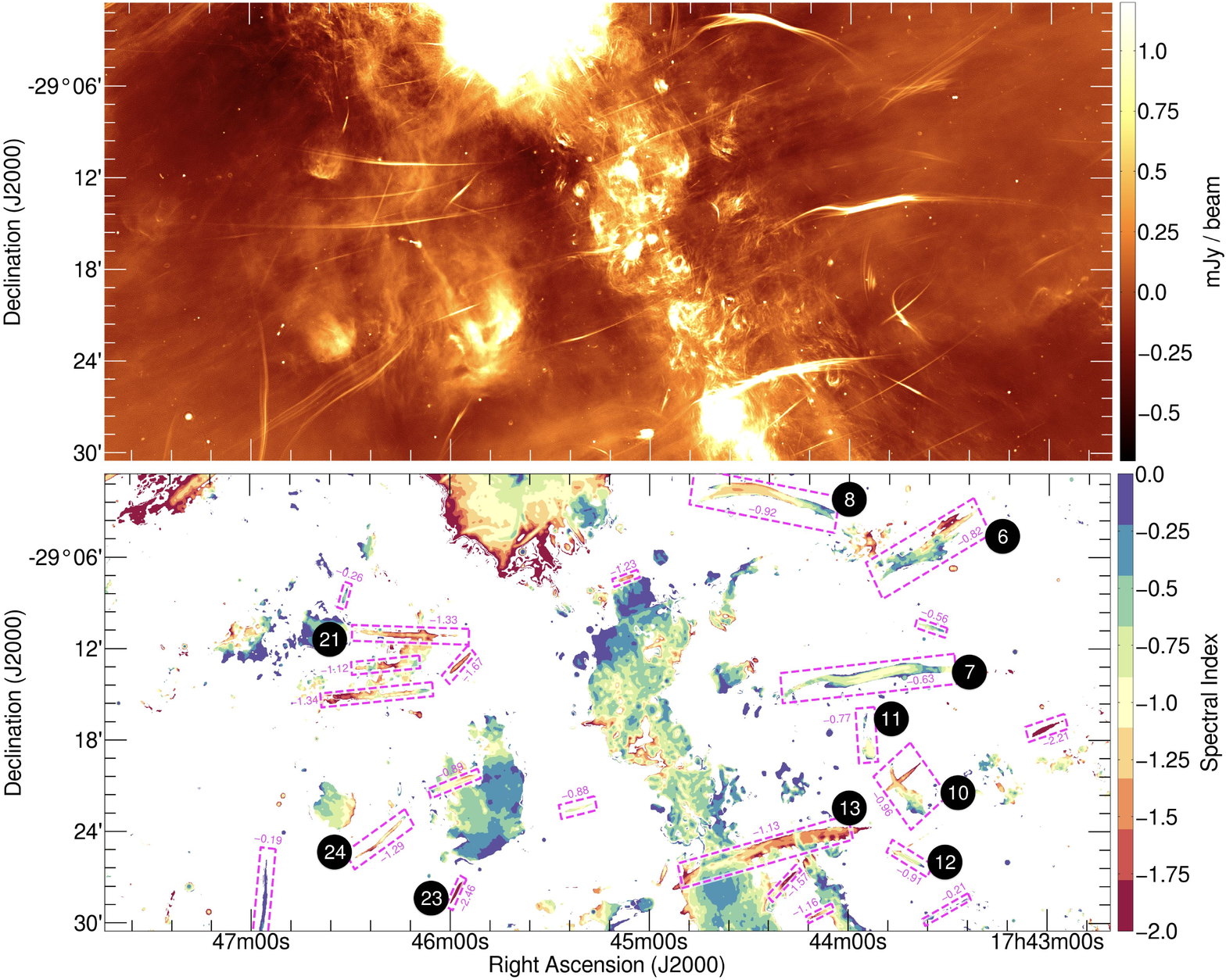}
\caption{Numerous non-thermal radio filaments spanning the Galactic plane between Sgr A and Sgr C, a region representing the western side of the collimated outflow from the Galactic center, as evidenced by X-ray and radio imaging. The upper panel shows the 1.28 GHz total intensity image at 4\dss\,resolution with a linear pixel stretch, and the lower panel shows the corresponding spectral index image, blanked in regions where the Stokes I brightness is below 200 $\mu$Jy beam$^{-1}$. The synchrotron-dominated emission from the Galactic plane runs diagonally across the image. The 23 rectangular magenta regions show the areas over which mean spectral index values are computed for each filament, with those values noted on the image. The numbers indicate panel numbers for filaments that also feature in Figure \ref{fig:filaments}.}
\label{fig:filaments_alpha}
\end{figure*}

Figure \ref{fig:filaments_alpha} shows numerous non-thermal radio filaments spanning the Galactic plane between the Sgr A and Sgr C regions. Total intensity and spectral index images are presented. The latter shows the synchrotron-dominated Galactic plane emission with numerous flatter spectrum source embedded within, as well as the broad range of spectral indices that the non-thermal filaments have. Spectral index gradients are visible along filament lengths, as have been previously seen in multi-band spectral index measurements \citep{larosa2001,law2008}. Mean spectral index values are calculated for 23 non-thermal filaments and filament complexes within the magenta regions visible in Figure \ref{fig:filaments_alpha}. The mean values of $\alpha$ are annotated on the figure close to or within each region. The overall mean value across all filaments measured here is $-$1.1~$\pm$~0.6 (1$\sigma$), consistent with synchrotron emission, and previously measured ranges \citep{law2008}, and also encompassing some filaments that are seen to have flatter spectra. The numbers shown on Figure \ref{fig:filaments_alpha} correspond to panel numbers on Figure \ref{fig:filaments} for any filaments that are also featured there.

For some of the filaments we can compare our in-band spectral index measurements to values derived from multi-band observations (\citealt{law2008}, see also \citealt{anantharamaiah1991,bally1989,yusef-zadeh2004}). Table \ref{tab:alphas} shows the mean in-band spectral indices for four such complexes as measured by MeerKAT, compared to the dual-frequency measurements from \citet{law2008}.

\begin{table}
\centering
\caption{In-band spectral indices for selected non-thermal filaments or filament complexes as measured by MeerKAT, compared to some dual-band measurements from \citet{law2008}. The MeerKAT ID refers to the panel number in Figures \ref{fig:filaments} and \ref{fig:filaments_alpha}, and the VLA ID refers to the identifications used in the aforementioned reference. Refer to Section \ref{sec:filaments} for details.}
\begin{tabular}{lllll} \hline \hline
ID$_{\mathrm{MeerKAT}}$ & $\alpha_{\mathrm{MeerKAT}}$ & ID$_{\mathrm{VLA}}$ & $\alpha^{6\mathrm{cm}}_{20\mathrm{cm}}$  \\ \hline
6 & $-$0.82 & N11a, N11b & $-$0.2 \\
7 & $-$0.63 & C3 & $-$0.8   \\ 
8 & $-$0.92 & N8 & $-$1.3 to $-$0.9 \\
10 & $-$0.96 & C6, C7 & $-$0.1 to $-$0.7 \\ \hline
\end{tabular}
\label{tab:alphas}
\end{table}

Some discrepancies are clearly present in the measurements listed in Table \ref{tab:alphas}, the most obvious explanation for which may be the differing angular resolutions between the two data sets. Addtionally, an unavoidable limitation of in-band spectral index measurements is that the resulting values have a stronger noise contribution than more widely separated dual-band measurements for a given signal to noise ratio, although this may be compensated for somewhat by the increased depth of the MeerKAT observations over the 6 and 20~cm VLA measurements primarily used to determine the above multi-band values. The differing methods used may also be a factor, namely our use of mean values determined within a rectangular aperture (containing thresholded data), versus the measurement of $\alpha$ in slices across the filament as employed by \citet{law2008}. The latter approch was designed to remove contributions from diffuse, background emission, and this may be contaminating our in-band measurements, particularly for those filaments closer to the plane and those seen at lower signal to noise ratios. Additionally, we are measuring the spectral index of filamentary features that are much fainter than those detected previously, encompassing previously undetected structures, and so the comparison of the spectral indices is fundamentally not a like-for-like one. For example, the filament complex in panel 6 of Figure \ref{fig:filaments}, is seen as a pair of filaments referred to as N11a and N11b by \citet{law2008}. MeerKAT resolves this complex into at least ten individual filaments.

The MeerKAT measurements are clearly able to recover reliable synchrotron spectra for the brighter filament complexes, as well as resolve previously-seen spectral index gradients along the filament. A detailed, statistical analysis of the bulk properties of the non-thermal filament population will be presented in a companion paper (Yusef-Zadeh et al., \emph{submitted}).

\subsection{Non-thermal radio filaments and their physical association with the 430 pc bipolar radio bubbles}

The previously-published MeerKAT observations of this region showed a bubble of radio emission with relativistic particles emerging from the Galactic center, with an end to end extent of 430 pc centered about the Galactic disk \citep{heywood2019}. The bubble appears symmetric about the Galactic plane but offset from Sgr A$^{*}$ by 20 pc towards negative longitudes, where the gas density in the disk is lower. X-ray observations indicate superheated gas distributed within the radio bubble, pointing to an energetic outflow with high cosmic ray flux in the nucleus \citep{nakashima2019,ponti2019}. The bubble encompasses prominent radio continuum sources, such as the Radio Arc with its network of nonthermal filaments at $l\sim0.2^\circ$, and Sgr C at $l\sim-0.5^\circ$. The eastern and western edges of the bubble, also known as the Galactic center radio lobes, show a mixture of warm ionized gas, dust and synchrotron emission \citep{sofue1984,tsuboi1995,bland-hawthorn2003,law2008}. Low-resolution radio recombination line observations of warm ionized gas along the radio lobes show velocities ranging between 20 and $-$20~km~s$^{-1}$ \citep{law2009,alves2015,nagoshi2019}. 

A key issue that arises is the location of this low velocity ionized gas, and its possible association with the nonthermal radio lobes. Low velocity ionized gas is generally distributed along the line of sight, much closer to us than the Galactic center. However, there are three indications that suggest the low-velocity gas is located in the GC region, and is associated with the Galactic center radio lobes. One is the association of the southern half of the radio bubble and the X-ray-emitting thermal plasma, with its high column of neutral absorbing material \citep{heywood2019,ponti2019,ponti2021}. Such high column density absorbing gas is unlikely to be at a local distance. Furthermore, linearly polarized emission extending along the eastern lobe, including the Radio Arc and its northern and southern extensions, show high Faraday rotation measures \citep{inoue1984,tsuboi1995,seiradakis1985,yusef-zadeh1997}. Such high values are consistent with the material being close to the Galactic center. Lastly, H$_3^+$ studies show the presence of the absorption line from (3,3) and the complete absence of lines from (2,2) toward the Galactic center. These are strong signatures that the gas is located in the Galactic center \citep{oka2005,oka2020}. The H$_3^+$ spectral measurements show warm ($\sim$200~K), diffuse molecular ionized gas, with a density of $\sim100$ cm$^{-3}$ and low velocities. In addition, a large volume filling factor of H$_3^+$ implies that the CMZ is not as opaque as previously thought. All these measurements together suggest that low velocity, low density gas is pervasive, and is likely to be distributed in the Galactic center region.

The cosmic ray ionization rate derived from H$_3^+$ measurements is at least 2$-$3 orders of magnitudes higher than in the Galactic disk \citep{lepetit2016,oka2020}, and thus can provide high pressure that can drive cosmic-ray driven wind. In this picture, the origin of low velocity ionized gas and dust is explained as the consequence of coronal gas pushing the warm ionized gas mixed in with dust in the direction where the edges of the bubble (the lobes) are, at $l\sim0.2^{\circ}$ and $l\sim-0.5^{\circ}$.

The bulk of the material in the bubble as traced by mid-infrared dust emission \citep{bland-hawthorn2003} is at the Galactic center, and arises from the ram pressure of coronal gas, accelerating and sweeping up dust and gas, and forming a shell-like structure. In this picture, a layer of low-velocity ionized gas is predicted to be interior to the linearly polarized radio lobes. 

The MeerKAT images suggest that the magnetized filaments and the large-scale radio bubble are causally connected because the majority of the linear filaments are distributed within the radio bubble cavities, and are spatially correlated \citep{heywood2019}. One scenario to explain the origin of the filaments is the interaction of a cosmic-ray driven wind with stellar wind bubbles, creating magnetized cometary tails \citep{yusef-zadeh2019}. In this model, the compression of the ambient cosmic ray electron population produces the synchrotron emission from the filaments.

An alternative explanation for the source of the high cosmic ray pressure in the Galactic center region could also be the leakage of cosmic ray particles from the filaments themselves, which subsequently interact with low-density gas distributed throughout the Galactic center. For this scenario to explain the bipolar radio / X-ray features would require a high volume filling factor of filaments in the region, and a high mean magnetic field strength in the filaments. The large number of new filaments revealed by the MeerKAT imaging provides evidence for the former (Yusef-Zadeh et al., \emph{submitted}).

\begin{figure*}[ht!]
\centering
\includegraphics[width=\textwidth]{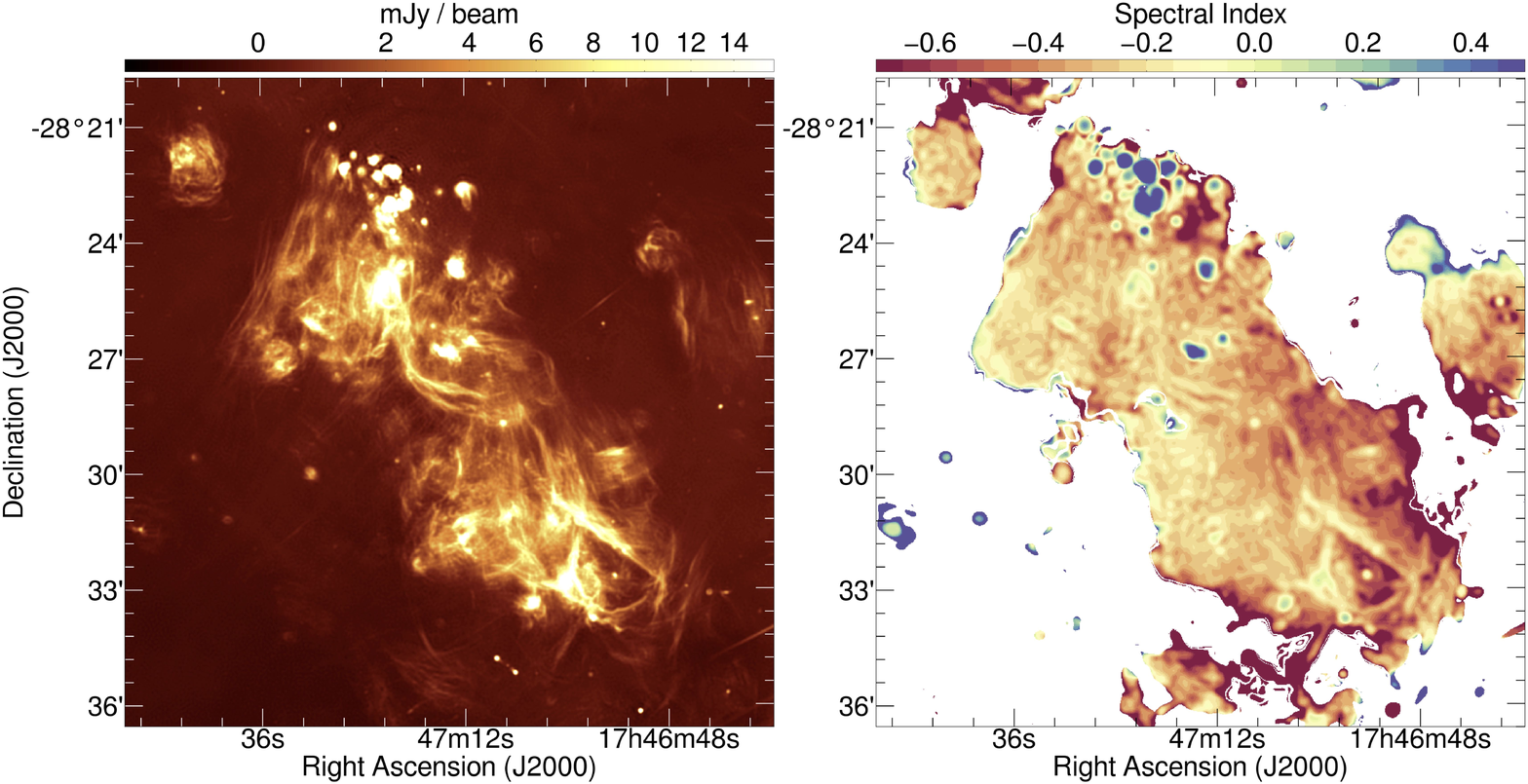}
\caption{The left hand panel shows 1.28 GHz radio emission from the Sagittarius B region (with a square-root pixel stretch) at an angular resolution of 4\dss. Sgr B is divided into two halves, with Sgr B1 to the south and Sgr B2 to the north. The right hand panel shows the corresponding spectral index image for this region, blanked below pixel values where the Stokes I brightness is below 300 $\mu$Jy~beam$^{-1}$. Sgr B2 is dominated by bright, compact, inverted spectrum radio features that signpost high-mass star-forming cores, as well as compact H\,\textsc{ii} regions. Sgr B1 consists mainly of filamentary and edge-brightened shell-like features with flatter radio spectra, although such features pervade the Sgr B complex as a whole. Refer to Section \ref{sec:sgrb} for further discussion.
\label{fig:sgrb}}
\end{figure*}

\subsection{Sagittarius B}
\label{sec:sgrb}

The Sagittarius B complex consists of two halves of roughly equal angular sizes: Sgr B1 (G0.5$-$0.0) to the south, and Sgr B2 (G0.7$-$0.0) to the north. The latter is a giant molecular cloud complex, containing many compact radio features that signpost extremely compact H\,\textsc{ii} regions and young stellar objects \citep{ginsburg2018}. Sgr B1 is thought to be the older half of the Sgr B complex, containing H\,\textsc{ii} regions that are more evolved, a molecular gas mass that is an order of magnitude lower than that of Sgr B2, and no evidence for on-going massive star formation \citep{henshaw2016,simpson2018,simpson2021}. The Sgr B2 region tends to be of particular interest as its mass, extent, and star formation rate suggest that the physical conditions within offer a local analog of the intense starburst regions found in ultraluminous infrared galaxies. Claims that the two regions are physically related are motivated by the presence of a common envelope of molecular material \citep{mehringer1992}, however alternative scenarios \citep{barnes2017} place the two halves of Sgr B on opposite sides of a common orbit around the Galactic center \citep{kruijssen2015}.

Figure \ref{fig:sgrb} shows the MeerKAT view of Sgr B at 1.28~GHz, with the total intensity image on the left panel, and the spectral index map of the corresponding region on the right. The Sgr B complex primarily consists of filamentary and edge-brightened, shell-like features, with Sgr B2 dominated by a number of bright, compact star-forming cores and H\,\textsc{ii} regions, clearly seen with inverted ($\alpha$$>$0.5) spectra. Sgr B1 is seen to have significantly fewer compact radio features, and those that are seen (primarily towards the southern edge) have flatter spectra than those in Sgr B2.

Previous radio imaging at 1.38 and 2.36 GHz with the ATCA \citep{jones2011} measured spectral indices for the compact and brighter ridge features, and concluded that the overall emission from Sgr B is dominated by thermal processes. The motivation for this work \citep[and previous work by][]{protheroe2008} was that synchrotron emission in Sgr B could be produced by secondary electrons and positrons that are produced when cosmic rays interact with the dense material in the clouds. Compact star forming cores aside, the MeerKAT imaging shows that the filamentary structures are generally seen with flatter spectral indices than their surrounds, but the general diffuse radio emission seen in the MeerKAT image may be a mixture of thermal and non-thermal processes as the spectral indices are generally flatter than those reported by \citet{jones2011}. Follow-up observations of this region (and many others) with MeerKAT's UHF (580--1015 MHz) and S-band receivers will be informative for modelling the spectra of such regions at high sensitivity and with high imaging fidelity. Coherent, thread-like radio features visible in Figure \ref{fig:sgrb} that span the entire Sgr B complex are also suggestive that Sgr B1 and B2 are related, if they are within the Sgr B complex and not a superposition.

\subsection{Sagittarius A and the Radio Arc Bubble}
\label{sec:sgra}

\begin{figure*}[ht!]
\centering
\includegraphics[width=0.95 \textwidth]{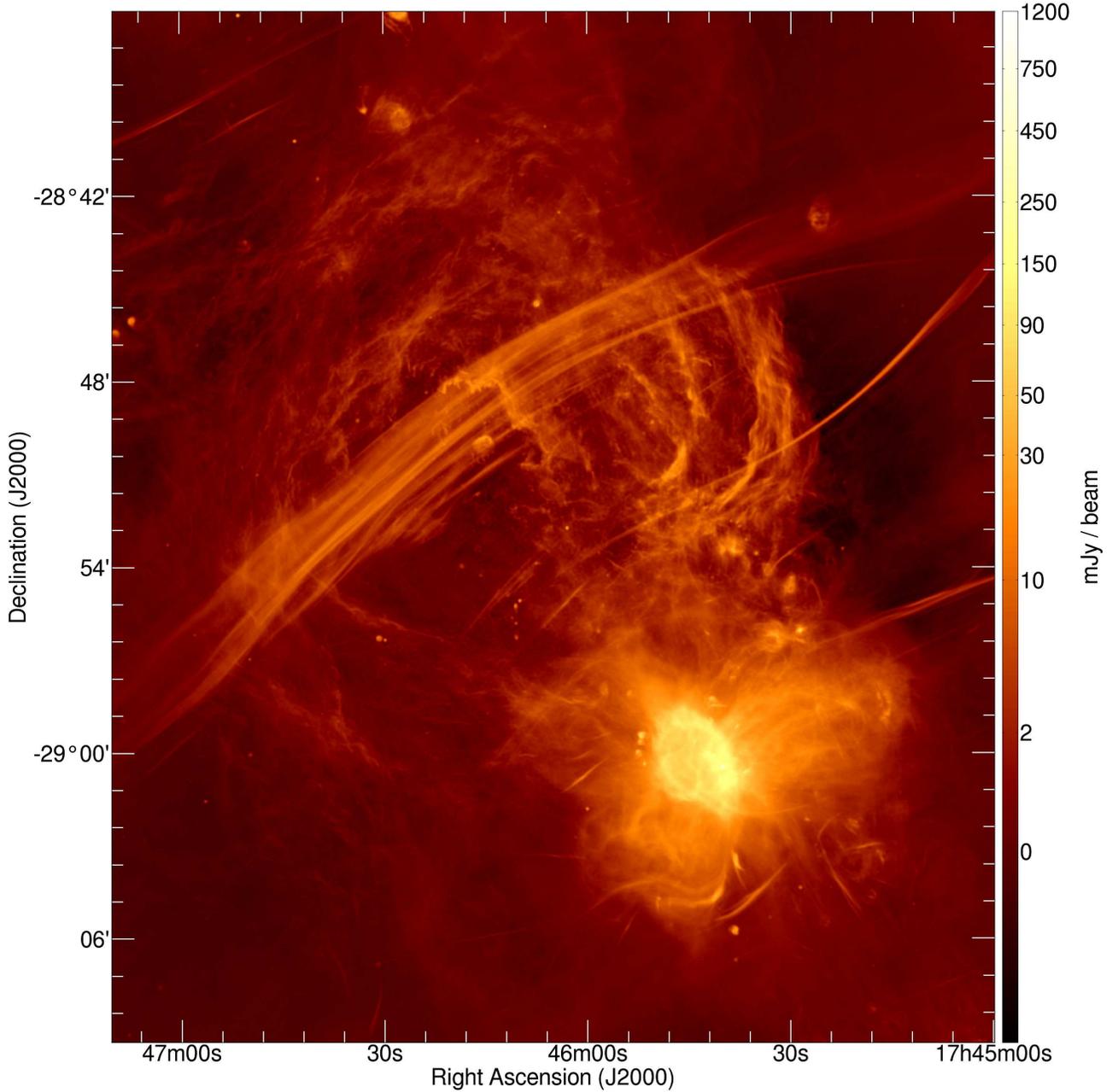}
\caption{The MeerKAT view of the Sgr A and Radio Arc Bubble regions, with an angular resolution of 4\dss. The image uses a logarithmic color stretch to capture the high dynamic range in the radio features visible. Many new filaments, compact sources, and low angular diameter shells can be seen in the figure, some closer views of which are shown in Figures \ref{fig:sickle} and \ref{fig:shells}. A desaturated image of the inner $\sim$30~pc region around Sgr A$^{*}$ is shown in Figure \ref{fig:sgra_alpha}, along with the corresponding spectral index image.
\label{fig:sgra}}
\end{figure*}

Figure \ref{fig:sgra} shows the 1.28 GHz radio emission from the region surrounding Sgr A$^{*}$ itself, and to the north, the radio emission associated with the quarter-degree diameter bubble structure generally known as the Radio Arc Bubble, as well as the Radio Arc filaments themselves. We discuss some of the features in this image in the subsections that follow.

\subsubsection{Interior of the Radio Arc Bubble}
\label{sec:interior}

\begin{figure*}[ht!]
\centering
\includegraphics[width= 0.95 \textwidth]{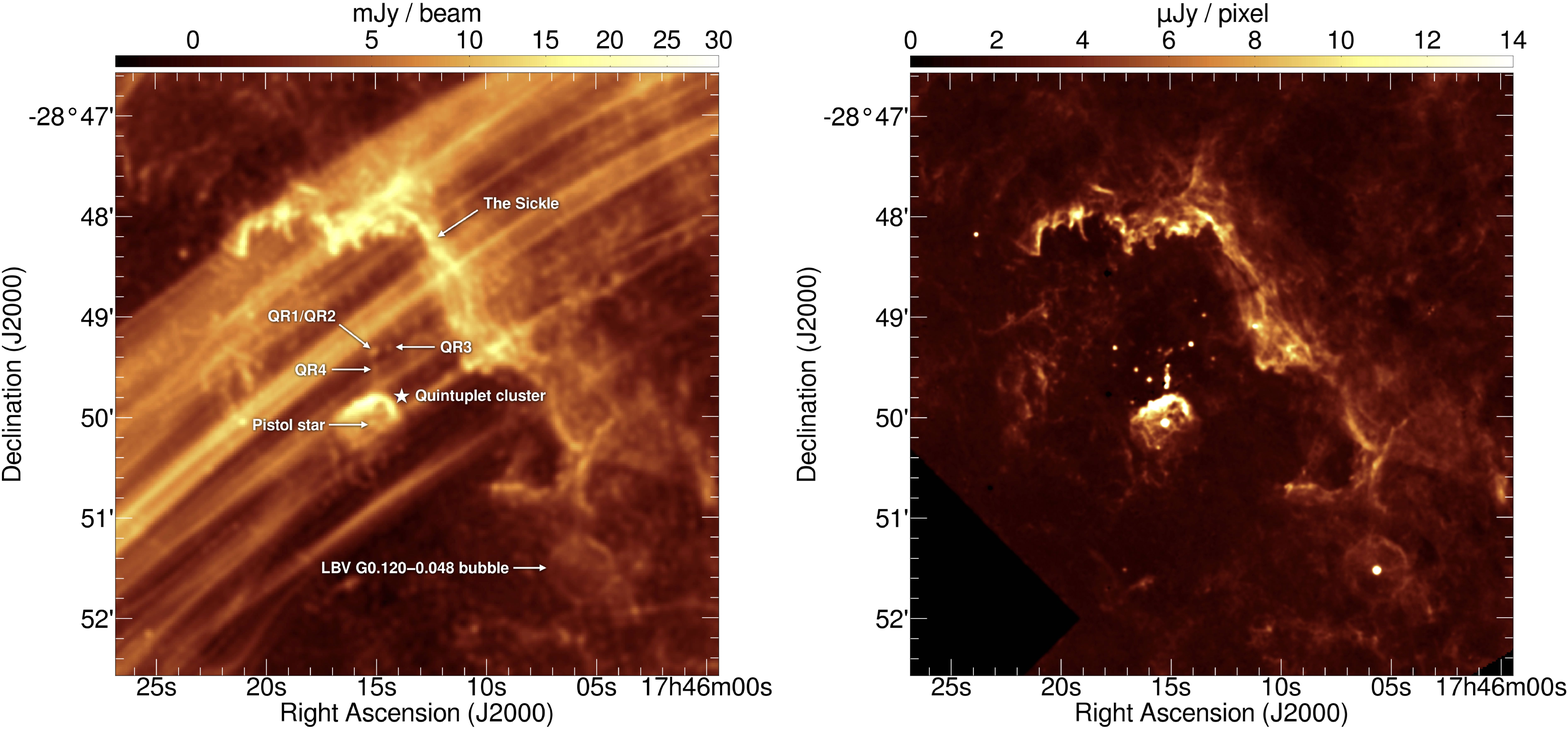}
\caption{MeerKAT image of the region surrounding the Quintuplet cluster at 1.28 GHz (left hand panel), with 4\dss\,angular resolution, and a square-root pixel stretch function as indicated on the color bar. The non-thermal filaments of the Radio Arc run diagonally across the image and dominate the scene, however we also see ``the Sickle" H\,\textsc{ii} region that is being ionized by the Quintuplet cluster, resolving pillar-like features in the ionization front in the radio imaging. The bubble surrounding the LBV star G0.120$-$0.048 can also be seen, a feature discovered in narrowband \emph{HST} imaging of the region \citep[right hand panel;][]{wang2010}. The well-known nebula surrouding the LBV Pistol star is also evident. The inverted spectrum continuum sources QR1--QR4 \citep{lang1999b} are detected at 1.28~GHz for the first time, although QR1 and QR2 are blended at the resolution of the MeerKAT data. Refer to Section \ref{sec:interior} for further details.
\label{fig:sickle}}
\end{figure*}

The Radio Arc Bubble is a shell-like structure 30 pc in diameter, traversed by the prominent Radio Arc filaments. The shell is thought to have been excavated by the cumulative pressure of supernovae from the stars within, and is seen at many wavelengths that can trace hot dust and ionized material. The interior of the Radio Arc Bubble is home to numerous massive, young star clusters, including the Arches \citep{espinoza2009} and Quintuplet \citep{liermann2009} clusters. MeerKAT imaging of the region surrounding the latter is presented in the left hand panel of Figure \ref{fig:sickle}. The Quintuplet cluster is the source of the ionizing photons that have produced ``the Sickle" H\,\textsc{ii} region at G0.18-0.04 \citep{figer1999}. The pillar-like features in the nebula's ionization front, as well as several low level features seen in \emph{Hubble} Space Telescope (\emph{HST}) narrowband imaging \citep[][right hand panel of Figure \ref{fig:sickle}]{wang2010} are visible in the MeerKAT imaging, although they are somewhat obscured by the non-thermal filaments that run across the image.

Some of the compact features in the image have been previously detected with higher angular resolution, higher frequency VLA observations \citep[sources QR1--QR4, as presented by][and indicated on Figure \ref{fig:sickle}]{lang1999b}. These sources are associated with the ionizing stellar winds from massive stars, have spectral indices (measured between 4.9 and 8.4 GHz) of between +0.5 and +0.8, and here we detect these faint sources at L-band frequencies for the first time.

We also detect the spherical nebula surrounding the luminous blue variable (LBV) star G0.120$-$0.048, thought to result from an outburst of material from the star itself \citep{mauerhan2010}, rather than from a wind interacting with the surrounding material. The 1.28 GHz detection is too faint to obtain a reliable in-band spectral index estimate, however this structure has previously been detected in a 24.5~GHz mosaic of the Sickle region by \citet{butterfield2018}. The well-known nebula surrounding the LBV Pistol star is also visible in Figure \ref{fig:sickle}.

\begin{figure}[h]
\centering
\includegraphics[width= \columnwidth]{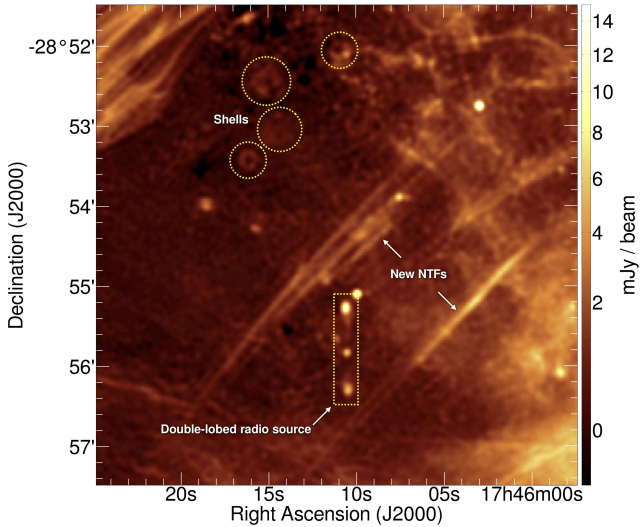}
\caption{This image shows new filaments and shells interior to the Radio Arc Bubble, as imaged by MeerKAT with an angular resolution of 4\dss. The image has a square-root stretch function as indicated on the color bar. A double-lobed radio source is also visible, as well as some low angular diameter shells, and new non-thermal filaments (NTFs). Refer to Section \ref{sec:interior} for details.}
\label{fig:shells}
\end{figure}

Figure \ref{fig:shells} shows a zoom of another region within the Radio Arc Bubble. In addition to the LBV shell, the MeerKAT imaging reveals several low angular diameter, low surface brightness shells as indicated on the figure, likely ionized H\,\textsc{ii} around the numerous massive stars in this region. 

There is a double-lobed radio source visible at the center of the lower third of Figure \ref{fig:shells}, with a lobe to lobe separation of 1\dmm. The core of this source has a cataloged X-ray counterpart at J2000 17\hhh46\mmm10\fs65 $-$28\ddd55\dmm50\farcs9 \citep{wang2006}, and the corresponding lobes have synchrotron spectra according to the MeerKAT spectral imaging. This source could be a background Fanaroff-Riley Type-II \citep[FR-II;][]{fanaroff1974} AGN, although MeerKAT is capable of resolving the expanding jets from Galactic X-ray binaries \cite[XRBs, e.g.][]{bright2020,carotenuto2021}. Examining the constituent images of the mosaic show that the lobes of this source appear to be stationary over a time scale of $\sim$1 month. However, this does not rule out a Galactic origin for this source, as the radio lobes could represent termination shocks rather than expanding ejecta. This would be comparable to the resolved jets of the Galactic microquasar 1E1740.7–2942 \citep[``the Great Annihilator";][]{mirabel1992} which is visible in Figure \ref{fig:mosaic_annotated}. The jets of 1E1740.7–2942 are seen to have synchrotron spectra in the spectral imaging, and are persistent.

Figure \ref{fig:shells} also shows some new filaments within the Radio Arc Bubble, running parallel to the primary filaments of the Radio Arc itself. The new filaments include another ``harp-like" structure, as indicated on the figure. We note that this was previously detected as a coherent polarised radio feature by \citet{reich2003}, and in total intensity by \citet{yusef-zadeh2004}, however both of these observations did not transversely resolve the structure into multiple individual filaments.

\begin{figure}[h]
\centering
\includegraphics[width= \columnwidth]{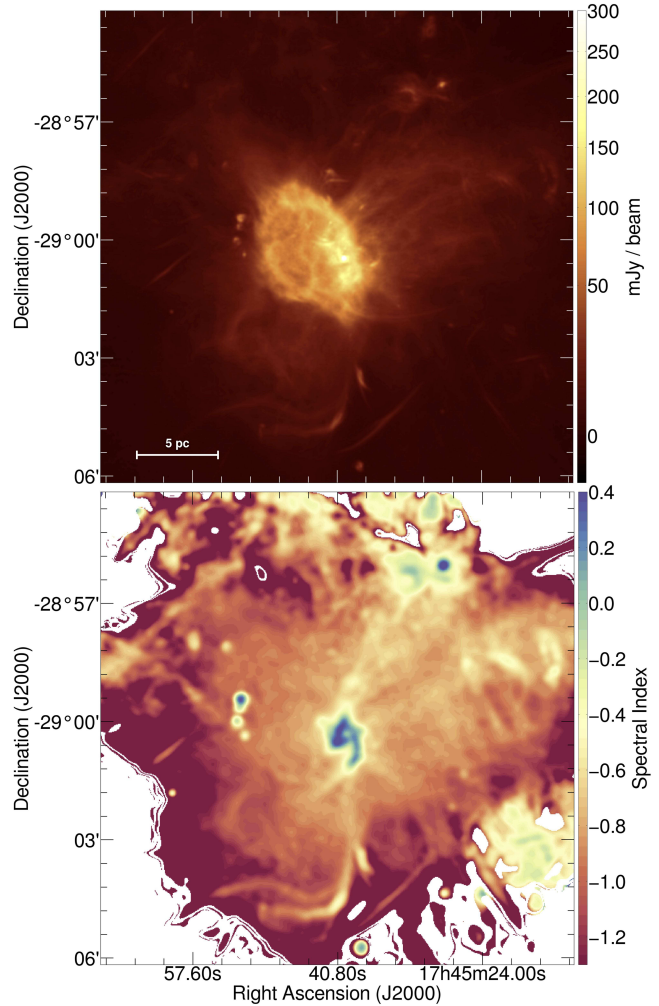}
\caption{A desaturated view of the 1.28 GHz radio emission from the Sgr A region is shown with 4\dss angular resolution in the upper panel (with a square-root pixel stretch), together with the spectral index image from the corresponding region in the lower panel. The bright Sgr A East shell-like structure is visible, at the convergence of a bipolar or hourglass shaped structure. The spectral index image is blanked for pixels where the Stokes I brightness is below 450 $\mu$Jy beam$^{-1}$, and clearly shows the thermal emission from the ionized mini-spiral, embedded in a synchrotron halo. The mini-spiral structure is apparently connected to two streams of material that flow through the center of the bipolar structure. Refer to Section \ref{sec:30pc} for details.}
\label{fig:sgra_alpha}
\end{figure}

\subsubsection{Accretion streams and outflows in the inner 30 pc of the Galaxy}
\label{sec:30pc}

Previous high resolution imaging of the Sgr A$^{*}$ region revealed an ionized gas feature known as the Galactic center mini-spiral \citep[e.g.][]{pedlar1989,scoville2003}. Observations tracing the kinematics of the gas along the arms of the mini-spiral \citep[e.g.][]{tsuboi2017} infer that the spiral arms are in Keplerian orbits surrounding Sgr A$^{*}$ \citep{hsieh2021}. Spectral imaging with MeerKAT's L-band (Figure \ref{fig:sgra_alpha}, lower panel) clearly reveals the spiral structure and the ionized nature of the gas via its inverted spectrum, embedded in the broader, diffuse synchrotron halo that surrounds Sgr A$^{*}$. Its connection to coherent, flatter spectrum features extending to the edge of the halo is visible in the spectral index image, a connection that is not obvious when looking at the total intensity images (Figure \ref{fig:sgra} and Figure \ref{fig:sgra_alpha}, upper panel) alone. These structures as a whole could represent the dominant accretion flow onto Sgr A$^{*}$, with the gas becoming ionised when it approches the inner $\sim$2 pc region where the mini-spiral is situated. A bipolar outflow from the inner $\sim$30 pc region has been inferred using high-resolution VLA observations at 5.5~GHz by \citet{zhao2016}. This study notes the correspondence between the radio emission to the northwest of Sgr A$^{*}$ and X-ray emission as detected by \emph{Chandra} \citep{markoff2010}. The MeerKAT imaging seems to show that the 5.5~GHz emission is part of a biconical or hourglass-shaped outflow, with the principal axis aligned with those of both the 430 pc radio bubbles, and the dominant non-thermal filaments in the surrounding region.

The bright, ring-like structure visible clearly in the upper panel of Figure \ref{fig:sgra_alpha} is not evident in the lower panel due to its synchrotron spectrum matching that of the broader Sgr A$^{*}$ halo. This structure, known as Sgr A East, is situated at the waist of the bipolar outflow, with the gas streams also converging towards Sgr A$^{*}$ through this region. The Sgr A East shell is typically interpreted to be a supernova remnant \citep{ekers1983,maeda2002,zhao2016}, however the inferred energetics are extreme, with a lower limit of 4~$\times$~10$^{52}$ erg \citep{mezger1989} being comparable to the derived energy budget of the 430 pc radio bubbles \citep{heywood2019}.

As is generally true for the broader mosaic, Figures \ref{fig:sgra} and \ref{fig:sgra_alpha} reveal many new filamentary structures. The filaments close to the Sgr A$^{*}$ halo show a much larger range of orientations compared to the broader mosaic, where filaments tend to be preferentially aligned perpendicular to the Galactic plane. This could be representative of a more complex magnetic field configuration in this region.

\section{Conclusion}
\label{sec:conclusion}

We have described the observations and data processing strategy for a new survey of the Galactic center at a central frequency of 1.28 GHz using the South African MeerKAT radio telescope. Twenty pointings have been processed, resulting in a mosaic that covers 6.5 square degrees of the region with 4\dss~angular resolution. The telescope's L-band receivers cover 856--1712 MHz, and we imaged the data in 16 sub-bands to produce an in-band spectral index mosaic of the region with an angular resolution of 8\dss. We used high (1\dss) resolution observations at 1--2 GHz from the VLA to verify the astrometry of the MeerKAT mosaic.

Some early science results from the survey are presented here, covering a variety of astrophysical phenomena on a range of scales. New discoveries in the field include potential radio supernova remnants, stellar bubbles driven by outbursts or winds, and numerous new non-thermal filaments. The MeerKAT imaging strongly suggests that the non-thermal radio filaments have a causal relationship with the 430 pc bipolar radio bubbles \citep{heywood2019}, and will facilitate future statistical studies of large numbers of these mysterious objects for the first time. The total intensity and spectral imaging also reveals the inflow and outflow processes at play in the inner few tens of parsecs of the Galaxy in unprecendeted detail, suggesting a bipolar or hourglass shaped outflow in this region, with its principal axis aligned perpendicular to the disk, parallel to that of the larger scale synchrotron bubbles.

Follow-up data products will include two catalogs derived from the image products we present here, namely a catalog of newly-discovered compact and point-like radio components (Rammala et al., \emph{in prep.}), and a catalog of filamentary structures extracted from high-pass filtered versions of the MeerKAT images (Yusef-Zadeh et al., \emph{submitted}). The production of polarimetrically calibrated radio images is also planned. We have demonstrated that the new radio view of the region afforded by MeerKAT represents a significant improvement over existing radio imaging of the field. The scientific potential of these products for studies of numerous phenomena in the region is high, both in their own right, as well as by combining the data we present here with the wealth of data available for the GC at other wavebands. The MeerKAT Galactic center observations are a SARAO public legacy project, and the imaging products presented here are made freely available with this article\footnote{\url{https://doi.org/10.48479/fyst-hj47}}. Raw visibility products are available from the SARAO data archive\footnote{\url{http://archive.sarao.ac.za}} under the project code SSV-20180505-FC-01.

\acknowledgments

The MeerKAT telescope is operated by the South African Radio Astronomy Observatory, which is a facility of the National Research Foundation, an agency of the Department of Science and Innovation. The authors acknowledge the contribution of all those who designed and built the MeerKAT instrument. The authors acknowledge the Centre for High Performance Computing (CHPC), South Africa, for providing computational resources to this research project. We acknowledge the use of the ilifu cloud computing facility – \url{www.ilifu.ac.za}, a partnership between the University of Cape Town, the University of the Western Cape, the University of Stellenbosch, Sol Plaatje University, the Cape Peninsula University of Technology and the South African Radio Astronomy Observatory. The ilifu facility is supported by contributions from the Inter-University Institute for Data Intensive Astronomy (IDIA – a partnership between the University of Cape Town, the University of Pretoria and the University of the Western Cape), the Computational Biology division at UCT and the Data Intensive Research Initiative of South Africa (DIRISA). The National Radio Astronomy Observatory is a facility of the National Science Foundation operated under cooperative agreement by Associated Universities, Inc. This research made use of Astropy,\footnote{\url{http://www.astropy.org}} a community-developed core Python package for Astronomy \citep{astropy:2013, astropy:2018}. This publication makes use of data products from the Wide-field Infrared Survey Explorer, which is a joint project of the University of California, Los Angeles, and the Jet Propulsion Laboratory/California Institute of Technology, funded by the National Aeronautics and Space Administration. This research has made use of NASA's Astrophysics Data System. This research made use of Montage, which is funded by the National Science Foundation under Grant Number ACI-1440620, and was previously funded by the National Aeronautics and Space Administration's Earth Science Technology Office, Computation Technologies Project, under Cooperative Agreement Number NCC5-626 between NASA and the California Institute of Technology. This work has made use of the Cube Analysis and Rendering Tool for Astronomy \citep[CARTA;][]{comrie2021}. This research has made use of Aladin Sky Atlas developed at CDS, Strasbourg Observatory, France \citep{bonnarel2000,boch2014}. IH acknowledges support from the UK Science and Technology Facilities Council [ST/N000919/1], and from the South African Radio Astronomy Observatory. FYZ is partially supported by the grant AST-0807400 from the the National Science Foundation. IH thanks Joe Bright, Rob Fender, and Aprajita Verma for useful discussions. Last but not least, the authors thank the anonymous referee for reviewing this article.

\vspace{5mm}
\facilities{MeerKAT, Karl G. Jansky Very Large Array}

\software{{\sc astropy} \citep{astropy:2013, astropy:2018}, {\sc carta} \citep{comrie2021}, {\sc casa} \citep{mcmullin2007}, {\sc eidos} \citep{asad2021}, {\sc matplotlib} \citep{hunter07}, {\sc montage}, {\sc oxkat} \citep{heywood2020}, {\sc pybdsf} \citep{mohan15}, {\sc pypher} \citep{boucaud16}, {\sc tricolour} \citep{hugo2021}, {\sc wsclean} \citep{offringa2014}}

\clearpage

\appendix
\section{Details of additional pointings}
\label{appendix}

The MeerKAT Galactic center survey consisted of 38 tracks in total, including the 20 that have been presented in this paper. The 18 tracks unused in this work are listed in Table \ref{tab:appendix} for completeness. Note that some fields were observed twice, so the number of unique pointing centers in the survey is 34. For all of these additional observations the correlator was configured to deliver 4,096 spectral channels with an integration time per visibility point of 8 seconds. All of these pointings have $\sim$10-hour track lengths, similar to those listed in Table \ref{tab:observations}. The additional unused pointings were outlier fields, including a series that extend above and below the Galactic plane to $|b|$$\sim$3$^{\circ}$. These encompass the pointings that formed the basis of the observations presented previously by \citet{heywood2019}, which are marked with an asterisk in Table \ref{tab:appendix}. 

\begin{table*}[h!]	
\begin{minipage}{175mm}
\centering
\caption{Summary of the additional MeerKAT observations of the Galactic center region that were not used in this paper. Field names highlighted with an asterisk formed the basis of the observations reported previously by \citet{heywood2019}. A plot of the locations of all of the pointings in the MeerKAT survey of the region can be found in Figure \ref{fig:all_ptgs}.}
\begin{tabular}{lllcccrr} \hline \hline
Date   & Block ID & N$_{\mathrm{ant}}$ & Name & \multicolumn{1}{c}{RA}              & \multicolumn{1}{c}{Dec}          & \multicolumn{1}{c}{$l$}     & \multicolumn{1}{c}{$b$}   \\ 
       &          &                    &      & \multicolumn{1}{c}{(hh:mm:ss.s)}    & \multicolumn{1}{c}{(dd:mm:ss.s)} & \multicolumn{1}{c}{(deg)}   & \multicolumn{1}{c}{(deg)} \\ \hline
2018-06-21 & 1529599329 & 61 & GCX17$^{*}$   & 17\hhh40\mmm36\fs7 &	$-$28\ddd39\dmm11\farcs4 & 359.664 & 1.082  \\
2018-06-22 & 1529686255	& 62 & GCX30$^{*}$   & 17\hhh50\mmm29\fs6 &	$-$29\ddd28\dmm45\farcs9 & 0.084   & $-$1.191 \\
2018-06-24 & 1529858765	& 62 & GCX33   & 17\hhh48\mmm54\fs5 &	$-$27\ddd25\dmm14\farcs4 & 1.672   & 0.165  \\
2018-06-25 & 1529944678	& 60 & GCXS30$^{*}$  & 17\hhh52\mmm16\fs4 &	$-$29\ddd39\dmm10\farcs5 & 0.133   & $-$1.613 \\
2018-07-02 & 1530548151	& 60 & GCX25   & 17\hhh44\mmm15\fs4 &	$-$27\ddd50\dmm48\farcs5 & 0.772   & 0.825  \\
2018-07-05 & 1530806454	& 62 & GCXN17$^{*}$  & 17\hhh38\mmm30\fs7 &	$-$28\ddd16\dmm00\farcs5 & 359.746 & 1.678  \\
2019-02-02 & 1549074680	& 61 & GCXN16  & 17\hhh37\mmm25\fs0 &	$-$28\ddd42\dmm00\farcs2 & 359.251 & 1.650   \\
2019-02-03 & 1549170075	& 61 & GCXN18  & 17\hhh39\mmm22\fs8 &	$-$27\ddd50\dmm13\farcs2 & 0.212   & 1.745  \\
2019-02-08 & 1549595596	& 61 & GCXNN17 & 17\hhh36\mmm09\fs2 &	$-$28\ddd01\dmm01\farcs4 & 359.678 & 2.251  \\
2019-03-16 & 1552779129 & 59 & GCXSS29 & 17\hhh53\mmm10\fs2 &   $-$30\ddd17\dmm38\farcs1 & 359.679 & $-$2.106 \\
2019-03-29 & 1553898724 & 60 & GCXSS30 & 17\hhh54\mmm05\fs6 &   $-$29\ddd50\dmm03\farcs1 & 0.177 & $-$2.045 \\
2019-03-30 & 1553985182 & 60 & GCXSS31 & 17\hhh54\mmm56\fs5 &   $-$29\ddd21\dmm45\farcs3 & 0.677 & $-$1.966 \\
2019-05-07 & 1557259251 & 59 & GCXSS29 & 17\hhh53\mmm10\fs2 &   $-$30\ddd17\dmm38\farcs1 & 359.679 & $-$2.106 \\
2019-05-12 & 1557694107 & 61 & GCXSS30 & 17\hhh54\mmm05\fs6 &   $-$29\ddd50\dmm03\farcs1 & 0.177 & $-$2.045 \\
2019-05-14 & 1557862253 & 57 & GCXSS31 & 17\hhh54\mmm56\fs5 &   $-$29\ddd21\dmm45\farcs3 & 0.677 & $-$1.966 \\
2019-07-05 & 1562353291 & 60 & TARGET & 17\hhh45\mmm15\fs5 & $-$28\ddd47\dmm35\farcs3 & 0.081 & 0.142 \\ 
2019-07-12 & 1562948237 & 61 & GCYN17 & 17\hhh34\mmm04\fs1 & $-$27\ddd38\dmm14\farcs6 & 359.749 & 2.844 \\
2019-07-13 & 1563036601 & 61 & GCYS30 & 17\hhh56\mmm30\fs0 & $-$30\ddd03\dmm08\farcs0 & 0.251 & $-$2.605 \\
\hline
\end{tabular}
\label{tab:appendix}
\end{minipage}
\end{table*}

\begin{figure*}
\centering
\includegraphics[width=0.95 \textwidth]{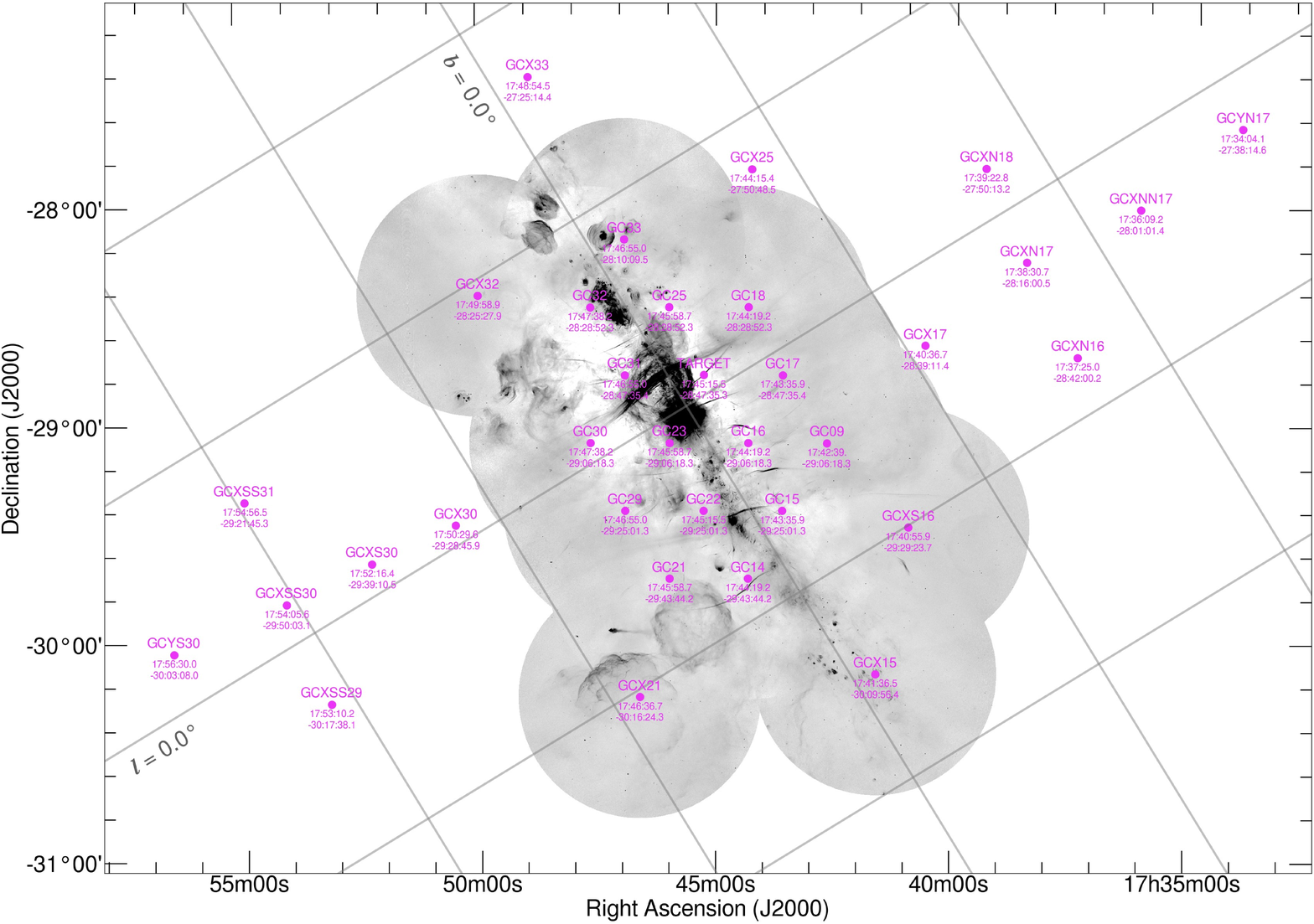}
\caption{The locations of all of the pointings that make up the MeerKAT survey of the Galactic center, shown in relation to the mosaic of the inner region presented in this paper. Further details can be found in Table \ref{tab:appendix}.
\label{fig:all_ptgs}}
\end{figure*}

{}

\end{document}